\documentclass[preprint,aps,prd,superscriptaddress,nofootinbib,longbibliography]{revtex4-2}
\usepackage{amsmath,amssymb,bm}
\usepackage{mathrsfs}
\usepackage{tikz}
\usetikzlibrary{shapes.geometric, arrows.meta, positioning}

\newcommand{\ed}{\eth}
\newcommand{\edb}{\bar{\eth}}
\newcommand{\ain}[1]{a^{\mathrm{in}}_{#1}}
\newcommand{\aind}[1]{a^{\dagger\,\mathrm{in}}_{#1}}
\newcommand{\aout}[1]{a^{\mathrm{out}}_{#1}}
\newcommand{\daout}[1]{\delta a^{\mathrm{out}}_{#1}}
\newcommand{\daoutd}[1]{\delta a^{\dagger\,\mathrm{out}}_{#1}}

\newcommand{\PV}{\operatorname{PV}}

\begin{document}

	\title{Quantum graviton scattering with definite helicities
in the null surface formulation. I:\\
Second-order scattering and the tree-level sector}

	\author{C.~N.~Kozameh}
	\email{carlos.kozameh@unc.edu.ar}
	\affiliation{FaMAF, Universidad Nacional de C\'ordoba,
		5000 C\'ordoba, Argentina}

	\author{G.~O.~Depaola}
	\affiliation{FaMAF, Universidad Nacional de C\'ordoba,
		5000 C\'ordoba, Argentina}

	\date{\today}

	\begin{abstract}
We develop a helicity-resolved description of quantum graviton scattering in
the Null Surface Formulation of asymptotically flat gravity.  In this
framework the basic dynamical data are the Bondi shear modes at null
infinity, and the spacetime metric is reconstructed from the associated
null-surface cut function $Z(x,\zeta)$.  The matching between
$\mathscr I^+$ and $\mathscr I^-$ is imposed at the level of the metric,
or equivalently on the directional derivative of the total cut function.
This leads to a decomposition
$$
Z_{\rm total}=Z_{\rm cut}+Z_{\rm cone},
$$
where $Z_{\rm cut}$ contains the free shear data and $Z_{\rm cone}$ is
the particular solution generated by the nonlinear cone source.  The
structure is directly analogous to the scalar relation between retarded and
advanced solutions, in which the difference of the free data is generated by
the Jordan--Pauli Green function and is therefore supported on shell.

Using the second-order NSF equations for $Z_2$ and $\Omega_2$, we derive
the helicity-resolved second-order Bondi shear and the corresponding
outgoing operators $\delta a^{\rm out}_{2,\pm}$.  These operators are
written entirely in terms of on-shell phase-space data at null infinity:
their kernels contain spin-weighted angular Green functions and their
momentum constraints are three-dimensional, with the outgoing energy fixed by
the positive-frequency on-shell condition.  The quadratic cone source
decomposes into normal-ordered $aa$, $a^\dagger a$, and
$a^\dagger a^\dagger$ sectors, so that different matrix elements select
different operator components of the same outgoing boundary operator.

As applications, we obtain the $(2\rightarrow1)$ gravitational tail amplitude
and the four-graviton $(2\rightarrow2)$ matrix elements.  The tail amplitude
is generated by the two-annihilation sector and selects the allowed
helicity combinations through the spin-weight structure of the NSF kernels.
For four-point scattering, the product of two outgoing boundary operators
reconstructs spatial momentum conservation directly.  For comparison with perturbative quantum gravity, we impose an additional
Poincar\'e reduction on the NSF matrix elements, replacing this
three-dimensional constraint by full four-momentum conservation.  Under
this reduction the second-order $2\rightarrow2$ matrix elements reproduce
the three Mandelstam exchange channels $s$, $t$, and $u$, with the
corresponding poles and helicity numerators arising from the angular Green
functions and helicity projections at null infinity.

The construction provides an intrinsically on-shell formulation of
perturbative graviton scattering in which no off-shell bulk propagators are
introduced.  Its natural observables are spectral-angular distributions on
the celestial sphere.  The Poincar\'e reduction is used only as an
additional comparison procedure with conventional flat-space amplitudes.
\end{abstract}

	\maketitle

\section{Introduction}
\label{sec:intro}
We study quantum graviton scattering in asymptotically flat gravity using
the Null Surface Formulation (NSF).  The formalism is well suited to this problem because its free gravitational
data are the radiative Bondi data at null infinity, while the interior
geometry is reconstructed from null surfaces determined by those data.  A brief historical and conceptual
review of the NSF, its relation to asymptotic quantization, and the
development from radiative phase space to nonlinear scattering is given in
Sec.~\ref{sec:NSF_review}.  Here we focus on the scattering problem and on
the quantum information extracted from the resulting relation between
incoming and outgoing radiative operators.

The goal of this paper is to develop the helicity-resolved quantum
scattering theory within the NSF at second order. The main
object is the nonlinear relation between the free incoming shear data and
the outgoing shear. At first order the matching between $\mathscr{I}^-$
and $\mathscr{I}^+$ gives the trivial identification of the free graviton
modes, up to the antipodal map. Nontrivial scattering begins at second
order, where the total cut function separates into a homogeneous cut part
and a particular cone part,
\begin{equation*}
Z_2=Z_{2,\mathrm{cut}}+Z_{2,\mathrm{cone}} .
\end{equation*}
The cut part contains the free second-order shear data, while the cone part
is generated by the quadratic source built from the first-order shear.

A central point of the construction is that the matching is imposed on the
metric reconstructed from the cuts, or equivalently on the directional
derivative of the total cut function. The reconstruction from free data on
$\mathscr{I}^+$ is an advanced solution for the metric, while the
reconstruction from free data on $\mathscr{I}^-$ is a retarded solution.
This terminology refers to the Green-function reconstruction of the
spacetime field, not to the names of the null coordinates $u$ and $v$.
The situation is directly analogous to a scalar field satisfying
\begin{equation*}
\Box\phi=J(\phi),
\end{equation*}
where the same solution may be written either in terms of future free data
and an advanced Green function, or in terms of past free data and a retarded
Green function. Equating the two representations gives a Jordan--Pauli
relation: the difference between free outgoing and incoming data is generated
by $G_{\rm ret}-G_{\rm adv}$, which is supported on shell. In the NSF the
analogous role is played by the matched cone contribution
$Z_{2,\mathrm{cone}}^+ + \widehat Z_{2,\mathrm{cone}}^-$.

Using the second-order NSF equations for $Z_2$ and $\Omega_2$, we derive
the helicity-resolved outgoing shear and extract the corresponding
second-order boundary operators
\begin{equation*}
\delta a^{\rm out}_{2,\pm}.
\end{equation*}
These operators are written entirely in terms of on-shell phase-space data at
null infinity. Their kernels contain spin-weighted angular Green functions,
helicity projectors, and the quadratic NSF source. Their momentum
constraints are three-dimensional, while the outgoing energy is fixed by the
positive-frequency on-shell condition. Thus the usual four-dimensional
Poincar\'{e} delta function is not inserted at the level of a single boundary
operator. It is imposed only in the Poincar\'{e}-reduced sector used for
comparison with conventional scattering.

The quadratic cone source has a normal-ordered sector decomposition. It
contains $aa$, $a^\dagger a$, and $a^\dagger a^\dagger$ components,
and different physical processes correspond to different matrix elements of
the same outgoing operator. The $2\rightarrow1$ gravitational tail
amplitude is obtained from a single insertion of
$\delta a^{\rm out}_{2,\pm}$ and selects the two-annihilation sector. The
spin-weight structure of the angular kernels imposes the allowed helicity
combinations. In particular, the tail amplitude in the two-annihilation sum channel
is supported by mixed-helicity incoming pairs.

Four-graviton scattering is obtained from the product of two outgoing
boundary operators. Each operator separately carries only a spatial
momentum constraint. After Wick contraction with a two-particle incoming
state, the product reconstructs conservation of external spatial momentum.
The Poincar\'{e} reduction is then defined by replacing this spatial constraint
with conservation of the total four-momentum. In this reduction the
resulting helicity amplitudes reproduce the standard tree-level graviton
scattering structure: the Mandelstam poles arise from the angular
Green-function and cone-kernel denominators, while the helicity numerators
come from the spin-weighted projections and null-tetrad contractions.

The formalism also gives a natural description of angular flux observables.
Since the Bondi news is expressed in terms of the same helicity-resolved
operators, the BMS supermomentum flux becomes an operator-valued angular
distribution on $\mathscr{I}^+$. It therefore probes the unreduced
spectral-angular data before imposing Poincar\'{e} momentum conservation. In
this sense scattering amplitudes, angular energy flux, and BMS
supermomentum are different projections of the same null-infinity operator
framework.

The paper is organized as follows. Section~\ref{sec:NSF_review} gives a
brief historical and conceptual review of the NSF, its relation to
asymptotic quantization and radiative phase space, and the developments that
led from the linearized quantum theory to nonlinear classical and quantum
scattering. In Sec.~\ref{sec:shear} we then introduce the helicity expansion
of the Bondi shear, formulate the metric matching in terms of the total cut
function, and derive the second-order outgoing operators from the NSF cone
source. We also explain the analogy with the Jordan--Pauli Green function
for a scalar field. In the following sections we apply the outgoing
operators to the $2\rightarrow1$ tail process and to four-graviton
$2\rightarrow2$ matrix elements. We then discuss BMS supermomentum flux and
its Bondi-energy content, the relation of the difference-frequency sector to
gravitational memory, and the Poincar\'{e} reduction used to compare the
four-graviton matrix elements with standard tree-level graviton amplitudes.
A subsequent section places the construction in relation with modern
asymptotic scattering, including BMS symmetries, soft-graviton methods,
gravitational memory, and on-shell amplitude techniques. We close with a
discussion of the resulting on-shell boundary picture of perturbative
graviton scattering.
\section{A Brief Review of the Null Surface Formulation of General Relativity}
\label{sec:NSF_review}

The Null Surface Formulation (NSF) provides an alternative description of
general relativity in which the conformal geometry of spacetime is encoded in
families of null hypersurfaces rather than being described primarily in terms
of the spacetime metric.  The basic observation is that the causal structure
of a Lorentzian spacetime is completely determined by its null cones.  One may
therefore reverse the usual point of view and regard a family of null
hypersurfaces as the fundamental geometrical object, reconstructing the
conformal metric from their structure.  The formulation was developed in a
series of works in which the vacuum Einstein equations were rewritten as
equations for null surfaces and a conformal factor
\cite{FKN95a,FKN95b,FKN95c}.

The central variable of the NSF is a scalar function
\begin{equation}
 Z=Z(x^a;\zeta,\bar\zeta),
 \label{eq:NSFreview_Z}
\end{equation}
defined on the sphere bundle of spacetime.  Here $x^a$ denotes a spacetime
point, while $(\zeta,\bar\zeta)$ are stereographic coordinates on the
celestial sphere and label null directions.  For every fixed
$(\zeta,\bar\zeta)$, the level surfaces
\begin{equation}
 Z(x^a;\zeta,\bar\zeta)=u
 \label{eq:NSFreview_level}
\end{equation}
are required to be null.  Consequently,
\begin{equation}
 g^{ab}\partial_a Z\,\partial_b Z=0 .
 \label{eq:NSFreview_eikonal}
\end{equation}
Thus a sphere's worth of solutions of the eikonal equation carries the
conformal information of the spacetime.

For an asymptotically flat spacetime this construction acquires a direct
physical interpretation.  Fixing a spacetime point $x^a$ and allowing
$(\zeta,\bar\zeta)$ to range over the sphere, the function
\begin{equation}
 u=Z(x^a;\zeta,\bar\zeta)
\end{equation}
describes the cut of future null infinity $\mathscr I^+$ produced by the
future null cone of $x^a$.  Conversely, for fixed
$(u,\zeta,\bar\zeta)\in\mathscr I^+$, the same equation describes the null
hypersurface extending from that asymptotic point into the interior.  These
two interpretations are reciprocal: $Z$ simultaneously describes light-cone
cuts at null infinity and the corresponding null hypersurfaces in the bulk.

The essential NSF statement can be represented schematically as
\begin{equation}
 \left\{Z(x;\zeta,\bar\zeta)\right\}
 \quad\Longleftrightarrow\quad
 [g_{ab}(x)] ,
 \label{eq:NSFreview_equivalence}
\end{equation}
where $[g_{ab}]$ denotes the conformal class of the spacetime metric.  A
second scalar variable, the conformal factor, completes the reconstruction of
the physical metric.  Not every function $Z$ defines a conformal spacetime
geometry.  Compatibility of the two-parameter family of null hypersurfaces
with a unique conformal metric imposes the NSF metricity conditions.  A
convenient variable is
\begin{equation}
 \Lambda=\eth^2 Z,
 \qquad
 \bar\Lambda=\bar\eth^{\,2}Z ,
 \label{eq:NSFreview_Lambda}
\end{equation}
whose allowed functional dependence is restricted by the metricity
conditions.  Together with the equation for the conformal factor, these
conditions provide a reformulation of the vacuum Einstein equations
\cite{FKN95a,FKN95b,FKN95c,BKR16}.

For asymptotically flat vacuum spacetimes the free characteristic data are
the Bondi shear, or equivalently the Bondi news, at null infinity.  This makes
the NSF particularly well adapted to a scattering problem.  The asymptotic
data determine the null surfaces and hence the interior conformal geometry,
while the asymptotic behavior of the reconstructed geometry determines the
corresponding radiative data.  The scattering problem can therefore be
regarded as a relation between two reconstructions of a single spacetime: an
advanced reconstruction based on data at $\mathscr I^-$ and a retarded
reconstruction based on data at $\mathscr I^+$.

\subsection{Radiative phase space and asymptotic quantization}

The use of radiative data at null infinity naturally connects the NSF with
the asymptotic quantization program of Ashtekar and with the
Ashtekar--Streubel symplectic description of gravitational radiation
\cite{Ashtekar81,Ashtekar1987,AshtekarStreubel}.  In that framework the
Bondi radiative data form a phase space equipped with a natural symplectic
structure.  Upon quantization, the corresponding Poisson algebra becomes the
algebra of asymptotic graviton observables.  This construction provides the
kinematical quantum framework for the free radiative degrees of freedom at
null infinity.

It is useful, however, to distinguish this kinematical construction from the
dynamical problem addressed by the NSF scattering program.  Asymptotic
quantization determines the radiative phase space and its quantum algebra,
but it does not by itself determine the nonlinear map relating incoming data
at $\mathscr I^-$ to outgoing data at $\mathscr I^+$.  In operator language,
the missing dynamical information is precisely that required to determine a
transformation of the form
\begin{equation}
 a^{\rm out}_{\lambda}
 =
 S^\dagger a^{\rm in}_{\lambda}S .
 \label{eq:NSFreview_Srelation}
\end{equation}

An important step toward such a dynamical theory was taken in
Ref.~\cite{DominguezKozamehLudvigsen}, where the global phase-space structure
of radiative spacetimes was analyzed.  The starting point was the continuity
of the leading curvature datum $\Psi_2$ through spatial infinity $i^0$.
On a nonradiative early cut, we use the Bondi mass aspect
$m_B=-\operatorname{Re}(\Psi_2^0+\sigma\dot{\bar\sigma})$ and define
$\chi=m_B(-\infty)=-\operatorname{Re}\Psi_2^0(-\infty)$.
The vacuum Bianchi identity gives
\[
 \dot m_B=\operatorname{Re}(\eth^2\dot{\bar\sigma})
            -\dot\sigma\dot{\bar\sigma}.
\]
Integrating from the early to the late cut gives
\begin{equation}
m_B(+\infty)-\chi
=
\operatorname{Re}\!\left(
\eth^2[\bar\sigma(+\infty)-\bar\sigma(-\infty)]
\right)
-
\int_{-\infty}^{+\infty}|\dot\sigma|^2\,du .
\label{eq:NSFreview_supermomentum}
\end{equation}
In the frame with $\sigma(+\infty)=0$ and with $m_B(+\infty)=0$,
the left-hand side reduces to $-\chi$.
The common early mass aspect labels the corresponding radiative
phase-space sector.

Several closely related representatives of gravitational supermomentum can
be introduced at null infinity.  For the present discussion the
Ashtekar--Streubel construction is especially convenient because the charge
and flux are naturally smeared with an arbitrary supertranslation function
and are directly tied to the symplectic structure of the radiative phase
space \cite{AshtekarStreubel}.  The lowest spherical-harmonic component gives
the Bondi energy balance, while the higher harmonics resolve the angular
supermomentum flux.  Thus nontrivial radiative evolution, including Bondi
mass loss, takes place within the global phase-space structure.

The phase-space analysis of Ref.~\cite{DominguezKozamehLudvigsen} also led
naturally to distinct complex structures associated with incoming and
outgoing radiative sectors and hence to the question of whether they can be
related by a unitary transformation.  Such a transformation would provide
the graviton $S$ operator.  In this sense that work established a
phase-space setting in which a subsequent nonlinear scattering dynamics
could be formulated, without yet constructing that dynamics.

Whether the nonlinear quantum scattering map developed in the present
program preserves the full foliation of the radiative solution space, or the
corresponding supermomentum sectors, is a separate dynamical question that we
do not address here.  Its possible relation with quantum superselection
sectors will be considered elsewhere.

\subsection{From linearized quantum geometry to nonlinear scattering}

The first quantum implementation of the NSF was carried out in the
linearized theory around Minkowski spacetime
\cite{FKNRT97}.  In that regime the Bondi data describe free gravitons and
the corresponding asymptotic Fock space is the Fock space of noninteracting
radiative modes.  The Bondi shear was promoted to an operator using
asymptotic quantization, and the linearized NSF equations were then used to
construct operator-valued null surfaces, null geodesics, and spacetime-point
variables.

This construction established a principle that remains central here: the
asymptotic radiative operators can serve as the fundamental quantum
variables, while geometrical quantities in the interior are obtained as
operator-valued functionals of those data.  However, because the theory was
linearized, it described quantum geometry generated by free gravitons and
contained no nontrivial graviton--graviton interaction.  In particular, it
did not determine a nonlinear scattering map between $\mathscr I^-$ and
$\mathscr I^+$.

A decisive development occurred in Ref.~\cite{BKR16}, where the nonlinear
NSF field equations were written in the definitive form used in the
subsequent development of the theory.  Their nonlinear dependence is
rational and is organized in terms of a single fundamental denominator,
$q$.  Multiplication by an appropriate power of $q$ converts the system into
an equivalent polynomial form.  This structural property makes the nonlinear
NSF equations particularly manageable and permits a systematic perturbative
expansion around Minkowski spacetime.

Using these equations, Ref.~\cite{BKR23} developed the perturbative NSF
through second order and formulated the classical gravitational scattering
problem.  The advanced solution determined from free data at $\mathscr I^-$
and the retarded solution determined from data at $\mathscr I^+$ were
required to represent the same interior spacetime.  This matching condition
produces a nonlinear relation
\begin{equation}
 \sigma_{\rm out}^{+}
 =
 {\cal F}\!\left[\sigma_{\rm in}^{-}\right]
 \label{eq:NSFreview_classicalmap}
\end{equation}
between incoming and outgoing Bondi data.  This is the classical scattering
map underlying the present construction.

The present work constitutes the next step in this sequence.  The nonlinear
classical scattering relation is promoted to the operator level,
\begin{equation}
 a^{\rm out}_{\lambda}
 =
 {\cal F}_{\lambda}
 \!\left[a^{\rm in},a^{{\rm in}\dagger}\right],
 \label{eq:NSFreview_quantummap}
\end{equation}
and therefore introduces, for the first time in this program, nontrivial
quantum graviton interactions.  The resulting outgoing operators are
nonlinear functionals of the incoming creation and annihilation operators
and generate the helicity-resolved matrix elements studied below.

This construction should be contrasted with the linearized quantum NSF.  The
latter quantizes the geometry generated by free radiative modes,
\begin{equation}
 \sigma\longrightarrow\widehat Z[\widehat\sigma],
\end{equation}
whereas the nonlinear scattering construction determines how interacting
incoming modes generate outgoing ones,
\begin{equation}
 \widehat\sigma^-_{\rm in}
 \longrightarrow
 \widehat Z
 \longrightarrow
 \widehat\sigma^+_{\rm out}.
 \label{eq:NSFreview_chain}
\end{equation}
The bulk geometry is therefore not introduced through independent off-shell
quantum fields; it is reconstructed through the NSF equations from
asymptotic on-shell radiative operators.

\section{Bondi shear, NSF matching, and outgoing operators}
\label{sec:shear}

The radiative degrees of freedom of an asymptotically flat spacetime are
encoded in the Bondi shear at null infinity, in the framework introduced by
Bondi and Sachs~\cite{Bondi62,Sachs62}. Following Ashtekar's radiative phase-space quantization at null
infinity~\cite{Ashtekar81,Ashtekar1987}, we expand the free shear in
helicity modes as
\begin{align}
\sigma_1^+(u,\zeta)
&=
\int_0^\infty
\frac{d\omega}{2\pi}
\sqrt{2\pi G}
\left[
a^{\rm in}_{+}(\omega,\zeta)e^{-i\varpi u}
+
a^{{\rm in}\dagger}_{-}(\omega,\zeta)e^{+i\varpi u}
\right],
\label{eq:sigma1_plus}
\\
\bar\sigma_1^+(u,\zeta)
&=
\int_0^\infty
\frac{d\omega}{2\pi}
\sqrt{2\pi G}
\left[
a^{\rm in}_{-}(\omega,\zeta)e^{-i\varpi u}
+
a^{{\rm in}\dagger}_{+}(\omega,\zeta)e^{+i\varpi u}
\right].
\label{eq:sigmabar1_plus}
\end{align}
Thus $a^{\rm in}_{+}$ and $a^{\rm in}_{-}$ annihilate gravitons of
helicity $+2$ and $-2$, respectively. We use the convention
\begin{equation}
\left[
a^{\rm in}_{\lambda}(\vec k),
a^{{\rm in}\dagger}_{\lambda'}(\vec k')
\right]
=
2\omega_k\,
\delta_{\lambda\lambda'}\,
\delta^{(3)}(\vec k-\vec k'),
\qquad
k^0=\omega_k=|\vec k|.
\label{eq:ccr_main}
\end{equation}

We label each Cartesian oscillator by the energy $\omega=|\vec k|$.
For $u=x\cdot\ell$, $\ell^0=1/\sqrt2$, put $\varpi=\sqrt2\,\omega$
(and $\varpi_j=\sqrt2\,\omega_j$). Then $k^a=\varpi\ell^a$.
Relative to Part~III we use $a_\lambda=\sqrt2\,a_\lambda^{\rm III}$,
$d\mu=d^3k/(2\omega)$ and the spatial distribution $\delta^{(3)}$.
Thus $\int d\mu(k')\,2\omega\delta^{(3)}(\vec k-\vec k')f(k')=f(k)$.
With $N=\sqrt{2\pi G}$ and $\nu(\omega)=N/(\pi\omega)$,
$d\mu\,\nu=N\,d\omega\,d\Omega/(2\pi)$. The two helicity contractions give
\[
[\sigma(u,\Omega),\bar\sigma(u',\Omega')]
=-\frac{2iG}{\pi}\delta_{S^2}^{(2)}(\Omega,\Omega')
\int_0^\infty\frac{d\omega}{\omega}\sin[\sqrt2\,\omega(u-u')]
=-iG\,{\rm sgn}(u-u')\delta_{S^2}^{(2)}(\Omega,\Omega').
\]

At first order the NSF matching between $\mathscr{I}^-$ and
$\mathscr{I}^+$ gives the trivial antipodal identification of the free
radiative modes. Nontrivial scattering starts at second order. We write
\begin{align}
\sigma_{\rm out}^+(u,\zeta)
&=
\epsilon\,\sigma_1^+(u,\zeta)
+
\epsilon^2\,\sigma_2^+(u,\zeta)
+
O(\epsilon^3),
\\
\bar\sigma_{\rm out}^+(u,\zeta)
&=
\epsilon\,\bar\sigma_1^+(u,\zeta)
+
\epsilon^2\,\bar\sigma_2^+(u,\zeta)
+
O(\epsilon^3),
\label{eq:sigma_out_expansion_main}
\end{align}
or, equivalently,
\begin{equation}
a^{\rm out}_{\lambda}
=
a^{\rm in}_{\lambda}
+
\epsilon\,\delta a^{\rm out}_{2,\lambda}
+
O(\epsilon^2).
\label{eq:aout_expansion_main}
\end{equation}
The purpose of this section is to explain the geometric origin of the
operators $\delta a^{\rm out}_{2,\lambda}$.

The central object in the null-surface formulation is the cut function
$Z(x,\zeta)$. For each fixed point $x^a$ in the space time,  the intersection of the future null cone from $x^a$ with future null infinity $\mathscr{I^+}$, with Bondi coordinates $(u,\zeta)$, is described by the equation,

\begin{equation}
u=Z(x,\zeta).
\end{equation}

The same equation has a second interpretation, for fixed values of $(u,\zeta)$ the level surfaces $Z(x,\zeta)=const.$ define null hypersurfaces of the spacetime.

The spacetime metric is reconstructed from this
family of cuts. In particular, the contraction of the metric perturbation
with the null direction $\ell^a(\zeta)$ is related to the directional
derivative of the total cut function by
\begin{equation}
h_{ab}(x)\ell^a(\zeta)\ell^b(\zeta)
=
-2\,\ell^a(\zeta)\partial_a Z_{\rm total}(x,\zeta).
\label{eq:hll_Ztotal_main}
\end{equation}
Thus the future--past matching is imposed on the metric reconstructed from
$Z_{\rm total}$, or equivalently on the directional derivative
$\ell^a\partial_a Z_{\rm total}$.

There are two reconstructions of the same physical metric. One is obtained
from the advanced cuts associated with $\mathscr{I}^+$, and the other from
the retarded cuts associated with $\mathscr{I}^-$. We denote the
corresponding total cuts by $Z_{\rm total}^+$ and
$Z_{\rm total}^-$. After the antipodal identification of the angular
variables, metric matching requires
\begin{equation}
g^+_{ab}[Z_{\rm total}^+]
=
g^-_{ab}[\widehat Z_{\rm total}^-],
\label{eq:metric_matching_main}
\end{equation}
where the hat denotes the antipodal transformation. Here ``advanced'' and ``retarded'' refer to the Green-function
reconstruction of the metric from free data, not to the name of the null
coordinate.  Thus the free data on \(\mathscr I^+\), although written in
terms of the retarded time \(u\), determine an advanced reconstruction of
the spacetime field.  Conversely, the free data on \(\mathscr I^-\), written
in terms of the advanced time \(v\), determine a retarded reconstruction.
This is precisely analogous to the scalar case, where the same solution can
be represented either from future data using an advanced Green function or
from past data using a retarded Green function.

Contracting with the
future null direction and using Eq.~\eqref{eq:hll_Ztotal_main}, this becomes
\begin{equation}
\ell^{+a}\partial_a
\left(
Z_{\rm total}^+
+
\widehat Z_{\rm total}^-
\right)
=
0 .
\label{eq:Ztotal_directional_matching_main}
\end{equation}
This is the form of the matching condition used in the perturbative
construction.

The vacuum Einstein equations are imposed on the cut $Z$  and a conformal factor $\Omega$ giving three scalar equations on a six dimensional space~\cite{BKR16}. The equations are  and invariant under BMS supertraslations and polynomial in the main variables, Thus one can obtain a hierarchy where the n-th perturbative equations use previous order terms.Ref.~\cite{BKR23}. The second-order NSF hierarchy is governed by the cut equation for $Z_2$
and the radial equation for the conformal factor $\Omega_2$. With
$\Lambda_1=\ed^2 Z_1$, the relevant equations are
\begin{align}
\edb^2\ed^2 Z_2
&=
\ed^2\bar\sigma(Z_1,\zeta,\bar\zeta)
+
\edb^2\sigma(Z_1,\zeta,\bar\zeta)
-\Sigma_2^+(Z_0)
\nonumber\\
&\quad
-
2\int_r^\infty
\left(
\ed\edb\,\Omega_2
+
\eta^{ab}\partial_a\Lambda_1\partial_b\bar\Lambda_1
\right)dr ,
\nonumber\\
[m_{TW}]_{-\infty}^{Z_0}
&=
2\big[\operatorname{Re}(\eth^2\bar\sigma)\big]_{-\infty}^{Z_0}
-\Sigma_2^+(Z_0),
\label{eq:NSF_Z2_main}
\\[6pt]
\partial_r^2
\left(
8\Omega_2
-
\partial_r\Lambda_1\,\partial_r\bar\Lambda_1
\right)
&=
\partial_r^2\Lambda_1\,
\partial_r^2\bar\Lambda_1 .
\label{eq:NSF_Omega2_radial_main}
\end{align}
These equations are valid for both the advanced construction from
$\mathscr{I}^+$ and the retarded construction from $\mathscr{I}^-$, with
the corresponding choice of null coordinate and antipodal angular
identification. We follow the conventions of Ref.~\cite{BKR23}.
The free boundary datum for the cut equation on $\mathscr I^+$ is the
Tamburino--Winicour mass aspect. We define it by
\[
m_{TW}
=m_B+\operatorname{Re}(\eth^2\bar\sigma)
=-\operatorname{Re}(\Psi_2^0+\sigma\dot{\bar\sigma}-\eth^2\bar\sigma).
\]
The notation $[F]_{-\infty}^{Z_0}$ means $F(Z_0)-F(-\infty)$, and
$\Sigma_2^+(Z_0)=\int_{-\infty}^{Z_0}|\dot\sigma_1^+|^2\,du$ is the
accumulated News flux at second order. The angular derivative acts at
fixed $u$ before evaluation on the cut. Integrating
$\partial_u m_{TW}=2\operatorname{Re}(\eth^2\dot{\bar\sigma})-|\dot\sigma|^2$
gives the boundary relation in Eq.~\eqref{eq:NSF_Z2_main}.

Equation~\eqref{eq:NSF_Omega2_radial_main} integrates to
\begin{equation}
8\Omega_2
=
\partial_r\Lambda_1\,\partial_r\bar\Lambda_1
+
\int_r^\infty dr'
\int_{r'}^\infty dr''\,
\partial_{r''}^{\,2}\Lambda_1\,
\partial_{r''}^{\,2}\bar\Lambda_1 .
\label{eq:NSF_Omega2_solution_main}
\end{equation}
Therefore the nonlinear source in the second-order cut equation is entirely
determined by the first-order cut $Z_1$, or equivalently by the linear
Bondi shear data.

It is useful to interpret the structure of these equations by analogy with a
scalar wave equation with source. For a massless scalar field satisfying
\begin{equation}
\Box\phi=J(\phi),
\label{eq:scalar_source_main}
\end{equation}
the same solution can be written either in retarded or advanced form,
\begin{align}
\phi(x)
&=
\phi_{\rm in}(x)
+
\int d^4x'\,
G_{\rm ret}(x,x')\,J(x'),
\label{eq:scalar_retarded_main}
\\
\phi(x)
&=
\phi_{\rm out}(x)
+
\int d^4x'\,
G_{\rm adv}(x,x')\,J(x').
\label{eq:scalar_advanced_main}
\end{align}
Equating the two representations gives
\begin{equation}
\phi_{\rm out}(x)-\phi_{\rm in}(x) = \int d^4x'\,\big[G_{\rm ret}(x,x')-G_{\rm adv}(x,x')\big]\,J(x').
\label{eq:scalar_JP_main}
\end{equation}
The difference $G_{\rm ret}-G_{\rm adv}$ is the Jordan--Pauli Green
function. It satisfies the homogeneous wave equation away from the source
and, in Fourier space, has support on the light cone. Thus the source
generates a difference between the free incoming and outgoing data through
on-shell light-cone propagation.

The NSF equations have the same causal structure, with the scalar field
$\phi$ replaced by the total null-surface function $Z_{\rm total}$. The
second-order solution is written as
\begin{equation}
Z_2^\pm
=
Z_{2,\mathrm{cut}}^\pm
+
Z_{2,\mathrm{cone}}^\pm .
\label{eq:Z2_split_main}
\end{equation}
The first term is the homogeneous cut part. Its boundary datum is the
second-order Tamburino--Winicour aspect, represented by the shear terms
and the signed flux in Eq.~\eqref{eq:NSF_Z2_main}. For the future construction,
\begin{equation}
\edb^2\ed^2 Z_{2,\mathrm{cut}}^+
=
\ed^2\bar\sigma_2^+
+
\edb^2\sigma_2^+-\Sigma_2^+(Z_0)
\label{eq:Z2cut_equation_main}
\end{equation}
The second term is the particular solution generated by the nonlinear cone
source,
\begin{equation}
\edb^2\ed^2 Z_{2,\mathrm{cone}}^+
=
-
2\int_r^\infty
\left(
\ed\edb\,\Omega_2^+
+
\eta^{ab}
\partial_a\Lambda_1^+\partial_b\bar\Lambda_1^+
\right)dr .
\label{eq:Z2cone_equation_main}
\end{equation}
Analogous equations hold for the retarded construction from
$\mathscr{I}^-$.

Substituting Eq.~\eqref{eq:Z2_split_main} into the metric matching condition, and fixing the incoming data as order 1, i.e. $\widehat Z_{2,\mathrm{cone}}^-=0$, we get
\eqref{eq:Ztotal_directional_matching_main} gives
\begin{equation}
\ell^{+a}\partial_a
\left(
Z_{2,\mathrm{cut}}^+
\right)
=
-
\ell^{+a}\partial_a
\left(
Z_{2,\mathrm{cone}}^+
+
\widehat Z_{2,\mathrm{cone}}^-
\right).
\label{eq:Z2_directional_matching_main}
\end{equation}
After fixing the homogeneous freedom in the cut normalization, this is
written equivalently as
\begin{equation}
Z_{2,\mathrm{cut}}^+ = -\left(Z_{2,\mathrm{cone}}^+ +
\widehat Z_{2,\mathrm{cone}}^-
\right).
\label{eq:Z2_matching_main}
\end{equation}
This is the NSF analogue of Eq.~\eqref{eq:scalar_JP_main}. The cone terms
on the right-hand side are known functionals of the first-order data. The
cut terms on the left-hand side contain the unknown second-order shear.
Therefore the incoming linear shear generates the second-order outgoing shear
through the cone part of the NSF solution.

In the quantum theory the first-order shear is expanded in creation and
annihilation operators. Substituting
Eqs.~\eqref{eq:sigma1_plus}--\eqref{eq:sigmabar1_plus} into the quadratic
cone source decomposes $Z_{2,\mathrm{cone}}$ into operator sectors,
\begin{equation}
Z_{2,\mathrm{cone}}^+
=
Z_{aa}^+
+
Z_{a^\dagger a}^+
+
Z_{a a^\dagger}^+
+
Z_{a^\dagger a^\dagger}^+ .
\label{eq:Zcone_sectors_main}
\end{equation}
After normal ordering, the relevant sectors are
\begin{equation}
aa,
\qquad
a^\dagger a,
\qquad
a^\dagger a^\dagger .
\label{eq:normal_ordered_sectors_main}
\end{equation}
The explicit evaluation of these sectors, including the angular kernels,
principal-value prescriptions, signs, and helicity structure, is given in
Appendix~\ref{app:Bcone}.

The Fourier phases of the normal-ordered sectors determine whether the
outgoing spatial momentum is carried by a difference channel or by a sum
channel.  The corresponding momenta are defined explicitly below, together
with the second-order outgoing operator.  In either case the outgoing
frequency is fixed by the positive-frequency on-shell condition
$p^0=|\vec p|$.  This is why the NSF operators are written with on-shell
measures and three-dimensional momentum delta functions, rather than with
off-shell four-dimensional propagators.

The spin-weight structure of the angular kernels selects the helicity
products that can appear in each sector. In particular, the
two-annihilation sector relevant for the $2\rightarrow1$ tail contains the
mixed-helicity pairs
\begin{equation}
a^{\rm in}_{+}(k_1)a^{\rm in}_{-}(k_2),
\qquad
a^{\rm in}_{-}(k_1)a^{\rm in}_{+}(k_2),
\label{eq:mixed_helicity_aa_main}
\end{equation}
while equal-helicity annihilation pairs do not contribute to that channel.
This selection rule is derived explicitly in Appendix~\ref{app:Bcone}.

Finally, the outgoing operators are read off from the positive-frequency
parts of the second-order shear. We write
\begin{align}
\sigma_2^+(u,\zeta)
&=
\int_0^\infty
\frac{d\omega}{2\pi}
\sqrt{2\pi G}
\left[
\delta a^{\rm out}_{2,+}(\omega,\zeta)e^{-i\varpi u}
+
\delta a^{{\rm out}\dagger}_{2,-}(\omega,\zeta)e^{+i\varpi u}
\right],
\label{eq:sigma2_mode_main}
\\
\bar\sigma_2^+(u,\zeta)
&=
\int_0^\infty
\frac{d\omega}{2\pi}
\sqrt{2\pi G}
\left[
\delta a^{\rm out}_{2,-}(\omega,\zeta)e^{-i\varpi u}
+
\delta a^{{\rm out}\dagger}_{2,+}(\omega,\zeta)e^{+i\varpi u}
\right].
\label{eq:sigmabar2_mode_main}
\end{align}
Comparing the solution of Eq.~\eqref{eq:Z2_matching_main} with
Eqs.~\eqref{eq:sigma2_mode_main}--\eqref{eq:sigmabar2_mode_main} defines
$\delta a^{\rm out}_{2,+}$ and $\delta a^{\rm out}_{2,-}$.

The resulting operators have the general form
\begin{equation}
\delta a^{\rm out}_{2,\lambda}(\vec p)
=
\int d\mu(k_1)\,d\mu(k_2)
\sum_C
\delta^{(3)}\!\left(\vec p-\vec K_C\right)
\mathcal K_C^{(\lambda)}(k_1,k_2;\hat p)\,
\mathcal O_C(k_1,k_2),
\label{eq:daout_general_main}
\end{equation}
with
\begin{equation}
d\mu(k)=\frac{d^3k}{2\omega_k},
\qquad
\omega_k=|\vec k|.
\label{eq:on_shell_measure_main}
\end{equation}
Here $C$ labels the sum and difference channels, $\mathcal O_C$ is one
of the normal-ordered operator products in
Eq.~\eqref{eq:deltaaout-B}, and
$\mathcal K_C^{(\lambda)}$ contains the spin-weighted angular kernels
generated by the NSF cone source and by the cut inversion.

The compact expression \eqref{eq:daout_general_main} displays only the
operator structure. The explicit second-order NSF result can be written
uniformly for both helicities as
\begin{equation}
	\delta a^{\rm out}_{2,\pm}(K)
	=
	-\frac{2w}{4\pi\sqrt2\,w\,\nu(w)}
	\sum_{i=\pm}
	\int d\mu_1\,d\mu_2
	\oint d^2z'\,
	\mathcal E_i^{(\pm)}(K;z')\,
	\mathcal Z_{i,\rm sym}^{(-)}\,
	\delta^{(3)}(\vec K-\vec K_i),
	\label{eq:daout2pm_explicit_main}
\end{equation}
where
\begin{equation}
	K^0=w=|\vec K|.
	\label{eq:K_on_shell_main}
\end{equation}
The two kinematic channels are
\begin{equation}
	\vec K_+=\vec k_1+\vec k_2,
	\qquad
	\vec K_-=\vec k_1-\vec k_2,
	\label{eq:Kpm_channels_main}
\end{equation}
and
\begin{equation}
	d\mu_i=\frac{d^3k_i}{2\omega_i},
	\qquad
	\omega_i=|\vec k_i|.
	\label{eq:dmu_i_main}
\end{equation}
The factors $\mathcal E_i^{(\pm)}(K;z')$ contain the helicity-dependent
projection and the corresponding angular Green-function factors.  The
coefficient $\mathcal Z_{i,\rm sym}^{(-)}$ is the Bose-symmetrized
negative-frequency coefficient in channel $i$, including the two input factors $\nu(\omega_1)\nu(\omega_2)$ as in Eq.~\eqref{eq:Lambda1-B}. The outgoing Fourier coefficient in $d^3K$ is $\nu(w)/(2w)$ and the News supplies $\sqrt2\,w$.  It contains the two
quadratic NSF contributions coming respectively from the conformal factor
$\Omega_2$ and from the $\Lambda_1\bar\Lambda_1$ term in the cut equation.

Equation~\eqref{eq:daout_general_main} gives the schematic operator
structure, whereas Eq.~\eqref{eq:daout2pm_explicit_main} gives the explicit
second-order NSF kernel. For a given matrix element only the sector with
the appropriate operator content contributes. For example, the
$2\rightarrow1$ tail amplitude selects the $aa$ projection
$\big[\mathcal Z_{+,\rm sym}^{(-)}\big]^{aa}$, whereas
four-point matrix elements may receive contributions from the mixed sectors
after the required Wick contractions.

The Fourier phases of the normal-ordered sectors determine whether the
outgoing spatial momentum is carried by a difference channel or by a sum
channel. This is implemented in Eq.~\eqref{eq:daout2pm_explicit_main} by
the three-dimensional delta functions
$\delta^{(3)}(\vec K-\vec K_i)$. The outgoing energy is then fixed on
shell by $K^0=w=|\vec K|$. Thus the NSF operators are written with
on-shell measures and three-dimensional momentum delta functions, rather
than with off-shell four-dimensional propagators.

Equation~\eqref{eq:daout2pm_explicit_main} is the operator input for the rest
of the paper. A single insertion gives the $2\rightarrow1$ tail amplitude,
while a product of two such operators gives the four-graviton
$2\rightarrow2$ amplitudes.

It is important to note that the delta functions appearing in
\(\delta a^{\mathrm{out}}_{2,\lambda}(\vec K)\) are three-dimensional Dirac
distributions. This is because the asymptotic graviton operator is defined
on null infinity and is labeled by three on-shell variables:
\[
\vec K=(\omega,\hat{K}),
\qquad
\omega=|\vec K|.
\]
Equivalently, the independent label is the spatial momentum \(\vec{K}=(\omega,z,\bar{z})\)
 on \(\mathscr{I}^+\). Thus, the Fourier extraction of
\(\delta a^{\mathrm{out}}_{2,\lambda}(p')\) produces distributions of the form
\[
\delta^{(3)}(\vec{p}\,'-\vec{K})
=
\frac{1}{\omega'^2}\,
\delta(\omega'-|\vec{K}|)\,
\delta^{(2)}(\hat{p}',\hat{K}),
\]
rather than a four-dimensional delta function. A four-dimensional
\(\delta^{(4)}\) would be appropriate for an off-shell bulk Fourier
transform or for the final covariant scattering amplitude after all external
on-shell matrix elements have been assembled. At the level of the
asymptotic operator \(\delta a^{\mathrm{out}}_{2,\lambda}(p')\), however, the
energy component is not an independent variable but is fixed by the null
on-shell condition.

In this sense, the frequency conjugate to the retarded time \(u\) should not
be regarded as an additional independent energy variable in the asymptotic
operator.  On \(\mathscr I^+\) it equals $\sqrt2$ times the radial energy $\omega^\prime$:
\[
\vec p'=\omega'\,\hat p',
\qquad
\omega'>0.
\]
Therefore the condition imposed by the Fourier extraction is a
three-dimensional momentum condition,
\[
\vec p'=\vec K,
\]
or equivalently
\[
\omega'\hat p'=\vec K.
\]
The delta function
\[
\delta^{(3)}(\vec p'-\vec K)
\]
simultaneously fixes the angular direction \(\hat p'\) and the radial
frequency \(\omega'\).  Thus the frequency appearing in the \(u\)-Fourier
transform is $\sqrt2$ times the radial variable in the three-momentum
conservation law.  No separate energy-conserving delta function is required
at the level of the asymptotic operator.




The outgoing map obtained above preserves the canonical commutation
relations through the order considered here.  Moreover, it is compatible
with the adjoint operation; in particular,
\begin{equation}
 \left(\delta a_{2}\right)^\dagger
 =
 \delta\!\left(a_{2}^\dagger\right),
 \label{eq:adjoint_da2}
\end{equation}
with the analogous property holding at lower orders.  The perturbative
transformation therefore defines, through second order, a canonical
$*$-map of the asymptotic bosonic operator algebra.  A general
reconstruction theorem for perturbative bosonic canonical transformations
then implies that such a map admits a unitary implementation through
second order: a self-adjoint generator can be reconstructed order by order
so that
\begin{equation}
 a^{\rm out}=S^\dagger a^{\rm in}S ,
\end{equation}
with $S$ unitary to the order under consideration.  We refer to
Ref.~\cite{KozamehCCR} for the general construction and proof.


\section{Three-graviton decay amplitude}
\label{sec:decay-amplitude}

\subsection{Decay matrix element and helicity selection}

We define the helicity-resolved $2\rightarrow1$ matrix element by
\begin{equation}
\mathcal M^{(+)}_{\lambda_1\lambda_2}(K_1,K_2;K)
=
\langle0|
\delta a^{\rm out}_{2,+}(K)
a_{\lambda_1}^\dagger(K_1)
a_{\lambda_2}^\dagger(K_2)
|0\rangle .
\label{eq:decay_matrix_element}
\end{equation}
Only the $aa$ sector of Eq.~\eqref{eq:daout2pm_explicit_main} contributes.
Using the canonical commutation relations and the Bose-symmetrized cone
kernel gives
\begin{equation}
\boxed{
\mathcal M^{(+)}_{\lambda_1\lambda_2}
=
\mathcal A^{(+)}_{\lambda_1\lambda_2}(K_1,K_2;K)\,
\delta^{(3)}(\vec K-\vec K_1-\vec K_2),
}
\label{eq:decay_amplitude_compact}
\end{equation}
where
\begin{equation}
\mathcal A^{(+)}_{\lambda_1\lambda_2}
=
-\frac{2w}{2\pi\sqrt2\,w\,\nu(w)}
\oint d^2z'\,
\mathcal E^{(+)}(K;K_1,K_2;z')\,
\mathcal Z^{\rm(sym)}_{\lambda_1\lambda_2}(K_1,K_2;z') .
\label{eq:decay_reduced_amplitude}
\end{equation}
Thus the Dirac distribution fixes the kinematical support, whereas the
reduced coefficient $\mathcal A^{(+)}_{\lambda_1\lambda_2}$ contains the
dynamics of the scattering process.

The annihilation sector contains only mixed-helicity pairs. Therefore
\begin{equation}
\boxed{
\mathcal M^{(+)}_{++}
=
\mathcal M^{(+)}_{--}
=
0,
\qquad
\mathcal M^{(+)}_{+-}
=
\mathcal M^{(+)}_{-+}.
}
\label{eq:decay_helicity_rule}
\end{equation}
This selection rule follows from the operator structure before evaluating
the angular integral.

\subsection{Kinematics and Bondi energy balance}

The spatial delta function in Eq.~\eqref{eq:decay_amplitude_compact} selects
\begin{equation}
\vec K=\vec K_1+\vec K_2 ,
\qquad
w=|\vec K_1+\vec K_2| ,
\label{eq:decay_kinematics}
\end{equation}
because the outgoing graviton is on shell.  Hence its final energy is
\begin{equation}
E_g^{\rm out}=w,
\end{equation}
whereas the two incoming gravitons carry
\begin{equation}
E_{\rm in}=\omega_1+\omega_2 .
\end{equation}
The triangle inequality therefore gives
\begin{equation}
\boxed{
\Delta E
\equiv
E_{\rm in}-E_g^{\rm out}
=
\omega_1+\omega_2-|\vec K_1+\vec K_2|
\ge0 ,
}
\label{eq:decay_energy_defect}
\end{equation}
with strict inequality for non-collinear incoming momenta and
$\Delta E=0$ only in the collinear limit.  Equation
\eqref{eq:decay_energy_defect} is a kinematical consequence of the
three-dimensional support; it is not a transition probability and does not
represent a failure of total energy conservation.

The Bondi mass-loss relation provides the appropriate asymptotic
framework for discussing the energy carried by the radiative field at
$\mathscr I^+$.  Since the relation between the News flux and the
interaction-generated outgoing operators requires some care, we derive it
explicitly in Sec.~\ref{sec:supermomentum}.  There the second-order shear is
differentiated to obtain the News, Parseval's identity is applied before any
momentum-space projection, and the resulting quadratic operator is evaluated
on the same two-graviton incoming state.  In particular, the physical
$2\rightarrow1$ projection remains supported on the sum channel
$\vec K=\vec K_1+\vec K_2$, so that its on-shell energy is
$|\vec K_1+\vec K_2|$.  The kinematical quantity $\Delta E$ in
Eq.~\eqref{eq:decay_energy_defect} should therefore be distinguished from
the probability-weighted energy carried by the interaction-generated
one-graviton tail.

\subsection{Collinear limit and wave-packet states}

The angular Green-function extraction is regular.  Near
$\ell'\rightarrow\ell_K$ one has
\begin{equation}
\ell_K\!\cdot\ell'=\mathcal O(\theta^2),
\qquad
m_K\!\cdot\ell'=\mathcal O(\theta),
\end{equation}
so that
\begin{equation}
G_{2,0'}(\hat K,z')
=
\frac{(m_K\!\cdot\ell')^2}{\ell_K\!\cdot\ell'}
=
\mathcal O(1),
\end{equation}
while the logarithmic part is locally integrable.  The possible singularity
instead comes from the affine factors
$\PV(\ell'\!\cdot K_+)^{-1}$.  For non-collinear incoming null momenta
$K_+=K_1+K_2$ is future timelike,
\begin{equation}
K_+^2=2K_1\!\cdot K_2>0 ,
\end{equation}
and therefore $\ell'\!\cdot K_+>0$ on the sphere.  The denominator can
vanish only in the collinear limit $K_+^2\rightarrow0$, which is also the
limit $\Delta E\rightarrow0$.

This plane-wave singularity is removed by using normalizable incoming
states,
\begin{equation}
A_{\lambda,f}^\dagger
=
\int d\mu(k)\,
f_\lambda(k)\,
a_\lambda^\dagger(k),
\end{equation}
with smooth profiles $f_\lambda$.  The collinear set has measure zero in
two-particle momentum space, so the smeared matrix element is a finite
distribution.  The singularity is therefore kinematical rather than
ultraviolet.


\section{Tail amplitudes as BMS supermomentum flux}
\label{sec:supermomentum}
The preceding section used the second-order outgoing operators to compute the three graviton
scattering amplitude. We now show that the same operators also give a
direct expression for an angular flux observable at null infinity, namely
the BMS supermomentum flux.

We use the Bondi mass aspect
\[
 m_B=-\operatorname{Re}(\Psi_2^0+\sigma\dot{\bar\sigma}).
\]
The real aspect used for the angular supermomentum in this section is
\begin{equation}
\Psi=\Psi_2^0+\sigma\dot{\bar\sigma}+\ed^2\bar\sigma .
\label{eq:real_mass_aspect}
\end{equation}
The asymptotic reality condition gives
\begin{equation}
\Psi=\bar\Psi
=-m_B+\operatorname{Re}(\ed^2\bar\sigma).
\label{eq:mass_aspect_reality}
\end{equation}
For a real supertranslation parameter $f(\zeta,\bar\zeta)$ the associated
charge
\begin{equation}
P_f=-\frac{1}{8\pi G}\oint d^2\zeta\,f(\zeta,\bar\zeta)\,\Psi
\label{eq:supermomentum_charge_f}
\end{equation}
is real. The function $f$ labels the BMS supertranslation generator: the
constant $f=\ell^0=1/\sqrt2$ gives the Bondi energy, the $\ell=1$ modes give the spatial
momentum, and the higher spherical harmonics give the genuine
supermomentum components.

Differentiating the relation between $\Psi$ and $m_B$ cancels the angular
term in $\dot m_B$ and gives the balance law
\begin{equation}
\dot\Psi = \dot\sigma\dot{\bar\sigma},
\label{eq:Bianchi_flux_balance}
\end{equation}
The sphere integral for the energy translation yields Bondi mass loss. We denote the outward flux by $\Delta P_f=P_f(-\infty)-P_f(+\infty)$. The radiative flux
associated with the same supertranslation parameter $f$ is therefore
\begin{equation}
\Delta P_f = \frac{1}{8\pi G} \int_{-\infty}^{+\infty}du \oint d^2\zeta \, f(\zeta,\bar\zeta) \, \dot\sigma^+(u,\zeta) \, \dot{\bar\sigma}^+(u,\zeta),
\label{eq:Pf_flux}
\end{equation}
where the normalization follows the Bondi-shear convention used throughout
this paper. Since $f$ is real and $\dot\sigma\dot{\bar\sigma}=|\dot\sigma|^2$,
the classical flux is real. In the quantum theory its representative must
therefore be a self-adjoint operator,
\begin{equation}
\Delta P_f^\dagger = \Delta P_f .
\label{eq:supermomentum_self_adjoint}
\end{equation}

\subsection{Bondi energy carried by the interaction-generated tail}
\label{subsec:bondi-tail-energy}

We now compute explicitly the Bondi energy carried by the radiative field
generated by the second-order interaction.  This calculation relates the
asymptotic News flux directly to the outgoing operator correction
$\delta a_{2,\mathrm{out}}$ entering the scattering matrix elements above.

The outgoing Bondi shear is expanded perturbatively as
\begin{equation}
\sigma_{\rm out}^{+}
=
\epsilon\,\sigma_1^{+}
+
\epsilon^2\,\sigma_2^{+}
+
O(\epsilon^3),
\label{eq:tail_sigma_expansion}
\end{equation}
where $\sigma_1^{+}$ contains the freely propagated radiative data and
$\sigma_2^{+}$ is the additional radiative field generated by the nonlinear
interaction.  We focus here on the energy carried by the latter
contribution.  Using the same helicity conventions as in
Sec.~\ref{sec:shear},
\begin{align}
\sigma_2^{+}(u,\zeta)
&=
\int_0^\infty \frac{d\omega}{2\pi}
\sqrt{2\pi G}
\left[
\delta a^{\rm out}_{2,+}(\omega,\zeta)e^{-i\varpi u}
+
\delta a^{{\rm out}\dagger}_{2,-}(\omega,\zeta)e^{+i\varpi u}
\right],
\label{eq:tail_sigma2_modes}
\\
\bar\sigma_2^{+}(u,\zeta)
&=
\int_0^\infty \frac{d\omega}{2\pi}
\sqrt{2\pi G}
\left[
\delta a^{\rm out}_{2,-}(\omega,\zeta)e^{-i\varpi u}
+
\delta a^{{\rm out}\dagger}_{2,+}(\omega,\zeta)e^{+i\varpi u}
\right].
\label{eq:tail_barsigma2_modes}
\end{align}
The associated News is
\begin{equation}
N_2=\dot\sigma_2^{+},
\qquad
\bar N_2=\dot{\bar\sigma}_2^{+},
\label{eq:tail_news2_def}
\end{equation}
and therefore
\begin{align}
N_2(u,\zeta)
&=
-i\int_0^\infty \frac{d\omega}{2\pi}
\sqrt{2\pi G}\,\varpi
\left[
\delta a^{\rm out}_{2,+}(\omega,\zeta)e^{-i\varpi u}
-
\delta a^{{\rm out}\dagger}_{2,-}(\omega,\zeta)e^{+i\varpi u}
\right],
\label{eq:tail_news2_modes}
\\
\bar N_2(u,\zeta)
&=
-i\int_0^\infty \frac{d\omega'}{2\pi}
\sqrt{2\pi G}\,\varpi'
\left[
\delta a^{\rm out}_{2,-}(\omega',\zeta)e^{-i\varpi' u}
-
\delta a^{{\rm out}\dagger}_{2,+}(\omega',\zeta)e^{+i\varpi' u}
\right].
\label{eq:tail_barnews2_modes}
\end{align}

Taking the time translation $f=\ell^0=1/\sqrt2$, the Bondi energy carried by this interaction-generated radiative field is
\begin{equation}
E_{\rm tail}
=
\frac{1}{8\sqrt2\,\pi G}
\int_{-\infty}^{+\infty}du
\oint d^2\zeta\,
N_2(u,\zeta)\bar N_2(u,\zeta).
\label{eq:Bondi_tail_sigma2}
\end{equation}
Substituting Eqs.~\eqref{eq:tail_news2_modes} and
\eqref{eq:tail_barnews2_modes} produces phases
$e^{\mp i(\varpi-\varpi')u}$ and
$e^{\mp i(\varpi+\varpi')u}$.  The retarded-time integration gives
\begin{equation}
\int_{-\infty}^{+\infty}du\,
e^{-i(\varpi-\varpi')u}
=
\frac{2\pi}{\sqrt2}\,\delta(\omega-\omega'),
\label{eq:tail_parseval_delta}
\end{equation}
whereas, for $\omega,\omega'>0$, the terms proportional to
$\delta(\omega+\omega')$ have no support away from the zero-frequency
endpoint.  Thus, for the finite-frequency process considered here, the
retarded-time integral selects the difference-frequency terms.  Explicitly,
\begin{align}
E_{\rm tail}
&=
\frac{1}{8\sqrt2\,\pi G}
\oint d^2\zeta
\int_0^\infty
\frac{d\omega\,d\omega'}{(2\pi)^2}
\,2\pi G\,\varpi\varpi'
\,(2\pi/\sqrt2)\delta(\omega-\omega')
\nonumber\\
&\quad\times
\left[
\delta a^{\rm out}_{2,+}(\omega,\zeta)
\delta a^{{\rm out}\dagger}_{2,+}(\omega',\zeta)
+
\delta a^{{\rm out}\dagger}_{2,-}(\omega,\zeta)
\delta a^{\rm out}_{2,-}(\omega',\zeta)
\right].
\label{eq:tail_parseval_explicit}
\end{align}
After carrying out the $\omega'$ integration and normal ordering, with the
vacuum contribution removed, one obtains
\begin{equation}
:\widehat E_{\rm tail}:
=
\frac{1}{8\pi}
\sum_{\lambda=\pm2}
\oint d^2\zeta
\int_0^\infty d\omega\,
\omega^2\,
\delta a^{{\rm out}\dagger}_{2,\lambda}(\omega,\zeta)
\delta a^{\rm out}_{2,\lambda}(\omega,\zeta).
\label{eq:Bondi_tail_parseval}
\end{equation}
Equation~\eqref{eq:Bondi_tail_parseval} is the explicit Parseval
representation of the Bondi flux carried by the interaction-generated
field.  The sum over $\lambda=\pm2$ accounts for the two physical graviton
helicities.

For positive-frequency modes,
\begin{equation}
\vec K=\omega\,\hat K(\zeta),
\qquad
d^3K=\omega^2\,d\omega\,d\Omega .
\label{eq:tail_on_shell_measure}
\end{equation}
The same Cartesian operator occurs in both representations.
Since $d^3K=\omega^2d\omega\,d\Omega$, the density in
Eq.~\eqref{eq:Bondi_tail_parseval} is integrated with $d^3K$.

For the incoming state
\begin{equation}
|\mathrm{in}\rangle
=
a^{{\rm in}\dagger}_{\lambda_1}(K_1)
a^{{\rm in}\dagger}_{\lambda_2}(K_2)|0\rangle ,
\label{eq:tail_two_graviton_in}
\end{equation}
the contribution of the one-graviton outgoing sector to the expectation
value of Eq.~\eqref{eq:Bondi_tail_parseval} is proportional to
\begin{equation}
\sum_{\lambda_3=\pm2}
\frac{1}{8\pi}\int d^3K\,
\left|
\langle0|
\delta a^{\rm out}_{2,\lambda_3}(K)
a^{{\rm in}\dagger}_{\lambda_1}(K_1)
a^{{\rm in}\dagger}_{\lambda_2}(K_2)
|0\rangle
\right|^2 ,
\label{eq:Bondi_tail_projection}
\end{equation}
which is the vacuum projection of
$\langle{\rm in}|\delta a_2^\dagger(K)\delta a_2(K)|{\rm in}\rangle$
relevant to the $2\rightarrow1$ final sector.

The matrix element in Eq.~\eqref{eq:Bondi_tail_projection} is precisely the
one evaluated in Sec.~\ref{sec:decay-amplitude}.  The two incoming creation
operators select the $aa$ sector of $\delta a_{2,\rm out}$, and hence
\begin{equation}
\langle0|
\delta a^{\rm out}_{2,\lambda_3}(K)
a^{{\rm in}\dagger}_{\lambda_1}(K_1)
a^{{\rm in}\dagger}_{\lambda_2}(K_2)
|0\rangle
=
\mathcal A_{\lambda_3;\lambda_1\lambda_2}
\,
\delta^{(3)}
(\vec K-\vec K_1-\vec K_2).
\label{eq:Bondi_tail_sum_channel}
\end{equation}
Consequently,
\begin{equation}
\vec K=\vec K_1+\vec K_2,
\qquad
\omega=|\vec K_1+\vec K_2|.
\label{eq:Bondi_tail_sum_energy}
\end{equation}
After the standard treatment of the square of the plane-wave delta
function, or equivalently for normalized wave packets, the energy carried
by this one-graviton tail therefore has the kinematic structure
\begin{equation}
\boxed{
\langle E_{\rm tail}\rangle_{2\to1}
\propto
\sum_{\lambda_3=\pm2}
\left|
\mathcal A_{\lambda_3;\lambda_1\lambda_2}(K_1,K_2)
\right|^2
\left.\frac{\omega}{2\omega}\right|_{\vec K=\vec K_1+\vec K_2} .
}
\label{eq:Bondi_tail_final_energy}
\end{equation}
The ratio displays the energy insertion $|\vec K_1+\vec K_2|$ and
the $1/(2|\vec K_1+\vec K_2|)$ from the outgoing state measure separately.
Their product is constant for nonzero outgoing momentum. The proportionality sign keeps explicit the standard state-normalization
factor associated with plane-wave transition probabilities.  The important
point is that the energy weight is fixed unambiguously by the sum-channel
support of the physical $2\rightarrow1$ matrix element.

It is useful to distinguish two appearances of a ``difference'' in this
calculation.  The factor $\delta(\omega-\omega')$ in
Eq.~\eqref{eq:tail_parseval_delta} is the frequency matching produced by
the quadratic News flux and Parseval's identity.  It does not imply that
the physical $2\rightarrow1$ state is supported at
$|\vec K_1-\vec K_2|$.  The latter support is fixed only after the outgoing
operator is projected on the specified external state, and for the
$2\rightarrow1$ process it is the $aa$ sum channel
$\vec K=\vec K_1+\vec K_2$.

\subsection{Angular supermomentum weighting}
\label{subsec:tail_supermomentum_weighting}

The same calculation immediately gives the angularly weighted BMS
observable.  Replacing the energy weight $\ell^0$ by a real
$f(\zeta,\bar\zeta)$ in Eq.~\eqref{eq:Pf_flux} inserts this weight in
Eq.~\eqref{eq:Bondi_tail_parseval}.  For the $2\rightarrow1$ projection,
the three-dimensional delta function in
Eq.~\eqref{eq:Bondi_tail_sum_channel} fixes the celestial direction to
$\widehat{K_1+K_2}$.  Accordingly, up to the same normalization convention,
\begin{equation}
\boxed{
(\Delta P_f)_{2\to1}
\propto
f\!\left(\widehat{K_1+K_2}\right)
\left.\frac{\omega}{2\omega}\right|_{\vec K=\vec K_1+\vec K_2}
\sum_{\lambda_3=\pm2}
\left|
\mathcal A_{\lambda_3;\lambda_1\lambda_2}
\right|^2 .
}
\label{eq:tail_supermomentum_sum_channel}
\end{equation}
Thus the BMS weight probes the celestial direction selected by the same
sum-channel matrix element; it does not alter the underlying
three-dimensional kinematic support.

\section{Connection to gravitational memory}
\label{sec:memory}
The gravitational memory effect~\cite{Christodoulou91,Strominger14}
corresponds to a permanent displacement of inertial detectors after the
passage of gravitational radiation. At null infinity it is encoded in the
net change of the Bondi shear,
\begin{equation}
\Delta\sigma(\zeta) = \sigma^+\big|_{u\to+\infty} - \sigma^+\big|_{u\to-\infty}.
\label{eq:memory_def}
\end{equation}
In the perturbative expansion used in this paper,
\begin{equation}
\sigma^+ = \epsilon \, \sigma_1^+ + \epsilon^2 \, \sigma_2^+ + O(\epsilon^3).
\label{eq:memory_sigma_expansion}
\end{equation}
The first-order contribution satisfies the linear antipodal matching
condition
\begin{equation}
\sigma_1^+(u,\zeta) + \bar\sigma_1^-(u,\hat\zeta) = 0 .
\label{eq:linear_matching_memory}
\end{equation}
Thus the scattering process considered here has no nonlinear memory at
linear order,
\begin{equation}
\Delta\sigma_1 = 0 .
\label{eq:linear_memory_zero}
\end{equation}
The leading nonlinear memory is therefore controlled by the second-order
shear $\sigma_2^+$.

\subsection{Memory as a zero-frequency projection of the difference sector}
The second-order shear is expressed in terms of the outgoing nonlinear
operators derived in Sec.~\ref{sec:shear},
\begin{equation}
\sigma_2^+(u,\zeta) = \int_0^\infty \frac{d\omega}{2\pi} \sqrt{2\pi G} \left[ \delta a^{\rm out}_{2,+}(\omega,\zeta)e^{-i\varpi u} + \delta a^{{\rm out}\dagger}_{2,-}(\omega,\zeta)e^{+i\varpi u} \right].
\label{eq:sigma2_memory_modes}
\end{equation}
Equivalently, using the on-shell spatial momentum variable
$\vec K=\omega\hat K(\zeta)$, the second-order outgoing operator has the
schematic support
\begin{equation}
\delta a^{\rm out}_{2,\lambda}(\vec K)
=
\sum_{i=\pm}
\int d\mu_1\,d\mu_2\,
\mathcal K^{(\lambda)}_i(k_1,k_2;\hat K)\,
\mathcal O_i(k_1,k_2)\,
\delta^{(3)}(\vec K-\vec K_i),
\label{eq:daout_memory_support}
\end{equation}
with the sum and difference momenta defined in
Eq.~\eqref{eq:Kpm_channels_main}.  The $2\rightarrow1$ amplitude discussed
in Sec.~\ref{sec:decay-amplitude} selects the $aa$ sum sector and is
therefore supported at $\vec K=\vec K_+$.  The zero-frequency memory
projection probes a different normal-ordered sector of the same
second-order outgoing operator.  In particular, the mixed
$a^\dagger a$ contribution carries the difference momentum $\vec K_-$ and
admits the soft limit
\begin{equation}
\omega=|\vec K_-|\longrightarrow0,
\qquad
\vec K_-=\vec k_1-\vec k_2\longrightarrow0 .
\label{eq:memory_zero_frequency_limit}
\end{equation}
Thus the nonlinear memory projection and the finite-frequency
$2\rightarrow1$ amplitude should not be identified as the same kinematic
channel.  Rather, they probe different normal-ordered sectors of the same
second-order asymptotic operator.

More explicitly, the second-order memory can be represented as the
zero-frequency projection
\begin{equation}
\Delta\sigma_2(\zeta) = \lim_{\omega\to0^+} \frac{-i\omega\sqrt{2\pi G}}{2} \left[ \delta a^{\rm out}_{2,+}(\omega,\zeta) - \delta a^{{\rm out}\dagger}_{2,-}(\omega,\zeta) \right]_{\rm diff}.
\label{eq:memory_soft_projection}
\end{equation}
Here $\Delta\sigma_2=\int du\,\dot\sigma_2$ is the zero Fourier mode of the
News. We use the symmetric limit of its two boundary values, assuming that
the smeared News admits this limit. For the transform $\int du\,e^{isu}$
these values are $-iN\omega\,\delta a_{2,+}$ and
$+iN\omega\,\delta a_{2,-}^\dagger$ at $s=\pm\sqrt2\,\omega$.
The subscript ``diff'' indicates that only the difference channel of
$\delta a^{\rm out}_{2,\pm}$ is retained. Using
Eq.~\eqref{eq:daout_memory_support}, this projection selects
\begin{equation}
\delta^{(3)}(\vec K-\vec K_-) \quad\longrightarrow\quad \delta^{(3)}(\vec K_-),
\label{eq:memory_delta3_zero}
\end{equation}
namely the equal-momentum configuration $\vec k_1=\vec k_2$.

\subsection{Quantum collinear singularity}
For exact plane-wave states, the zero-frequency projection is
distributional. Consider the two-graviton state
\begin{equation}
|\psi\rangle = a^{{\rm in}\dagger}_{+}(\vec k_1) \, a^{{\rm in}\dagger}_{-}(\vec k_2) |0\rangle .
\label{eq:memory_twograviton_state}
\end{equation}
The memory channel requires $\vec k_1=\vec k_2$. In that limit the
canonical commutator produces the formal factor
\begin{equation}
[ a^{\rm in}_{\lambda}(\vec k), a^{{\rm in}\dagger}_{\lambda}(\vec k) ] = 2\omega_k \, \delta^{(3)}(\vec0),
\label{eq:delta0_memory}
\end{equation}
with the normalization conventions used in this work. Thus the memory
matrix element is singular for infinitely sharp momentum eigenstates.

This singularity is not a new ultraviolet divergence. It is the
distributional singularity associated with projecting an exact plane wave
onto the zero-frequency, equal-momentum endpoint of the difference sector. The tail
has finite support at
\begin{equation}
\omega = |\vec k_1-\vec k_2|,
\label{eq:tail_frequency_memory}
\end{equation}
whereas memory corresponds to the limiting configuration $\omega\to0$.

\subsection{Wave-packet regularization and detector resolution}
For physical wave packets the collinear singularity is smeared. It is
important to distinguish two different bandwidths. The first is the
preparation bandwidth $\Delta_{\rm in}$, which specifies the spectral and
angular width of the incoming wave-packet state. This bandwidth regularizes
the matrix element because the second-order outgoing operator is quadratic
in the incoming modes. The second is the detector bandwidth
$\Delta_{\rm det}$, which specifies the finite frequency and angular
resolution with which the outgoing memory observable is measured at
$\mathscr I^+$. The former belongs to the state preparation; the latter
belongs to the definition of the measured observable. They need not be
identified, although in a simplified model one may choose them to be of the
same order.

Let
\begin{equation}
a^{{\rm in}\dagger}_{G,\lambda} = \int d\mu(k) \, G_\lambda(k) \, a^{{\rm in}\dagger}_{\lambda}(k),
\label{eq:smeared_creation_memory}
\end{equation}
with $G_\lambda(k)$ a smooth wave-packet profile of width $\Delta_{\rm in}$.
We consider the two-graviton state
\begin{equation}
|\psi_G\rangle = a^{{\rm in}\dagger}_{G,+} \, a^{{\rm in}\dagger}_{G,-} |0\rangle .
\label{eq:smeared_twograviton_state}
\end{equation}
The memory expectation value is computed from the same second-order
operator,
\begin{equation}
\langle\Delta\sigma_2(\zeta)\rangle_G = \langle\psi_G| \Delta\sigma_2(\zeta) |\psi_G\rangle ,
\label{eq:smeared_memory_expectation}
\end{equation}
where
\begin{equation}
\Delta\sigma_2(\zeta) = \lim_{\omega\to0^+} \frac{-i\omega\sqrt{2\pi G}}{2} \left[ \delta a^{\rm out}_{2,+}(\omega,\zeta) - \delta a^{{\rm out}\dagger}_{2,-}(\omega,\zeta) \right]_{\rm diff}.
\label{eq:memory_soft_projection_smeared}
\end{equation}
No independent smearing of the outgoing mode is required to make the matrix
element finite: the outgoing second-order mode is induced by the incoming
packet through the quadratic operator $\delta a^{\rm out}_{2,\lambda}$.

Indeed, the difference channel contains the support
\begin{equation}
\delta^{(3)}(\vec K-\vec k_1+\vec k_2).
\label{eq:memory_packet_delta_support}
\end{equation}
In the plane-wave limit the memory projection $\vec K\to0$ imposes
$\vec k_1=\vec k_2$ and produces the formal factor $\delta^{(3)}(\vec0)$.
For the wave-packet state \eqref{eq:smeared_twograviton_state}, the same
delta function gives instead a finite overlap integral,
\begin{equation}
\int d\mu(k_1) \, d\mu(k_2) \, G_+(k_1) \, G_-(k_2) \, \delta^{(3)}(\vec k_1-\vec k_2) \, \mathcal K_{\rm diff}(k_1,k_2;\zeta),
\label{eq:memory_wavepacket_overlap}
\end{equation}
where $\mathcal K_{\rm diff}$ denotes the difference-channel NSF kernel.
Thus the singular equal-momentum condition is replaced by the finite overlap
of the two incoming wave packets near $\vec k_1=\vec k_2$.

One may nevertheless include an additional smearing of the outgoing
observable to model the finite resolution of a detector at null infinity.
For a detector profile $D_{\Delta_{\rm det}}(\omega,\zeta;\Omega,Z)$, the
measured soft memory mode is
\begin{equation}
\Delta\sigma_{D,2}(Z) = \int_0^\infty d\omega \oint d^2\zeta \, D_{\Delta_{\rm det}}(\omega,\zeta;0,Z) \frac{-i\omega\sqrt{2\pi G}}{2} \left[ \delta a^{\rm out}_{2,+}(\omega,\zeta) - \delta a^{{\rm out}\dagger}_{2,-}(\omega,\zeta) \right]_{\rm diff}.
\label{eq:detector_smeared_memory}
\end{equation}
This detector smearing is conceptually distinct from the wave-packet
regularization of the incoming state. The incoming smearing makes the
matrix element finite, while the detector smearing specifies the frequency
and angular resolution with which the outgoing memory observable is
measured.

In the classical coherent-state limit, the profiles $G_\lambda(k)$ and
$D_{\Delta_{\rm det}}$ are not auxiliary regulators but physical profiles
of the incoming radiation and of the measurement process. The sharp
plane-wave expression is recovered only as a singular idealization in which
the packet and detector widths are taken to zero.

\subsection{Memory, tails, and BMS symmetry}
The NSF derivation exhibits a simple operator hierarchy.  The same
second-order outgoing operator $\delta a^{\rm out}_{2,\pm}$ contains
different normal-ordered sectors that are selected by different
observables.  The $2\rightarrow1$ matrix element selects the $aa$ sum
channel and is supported at
$\omega=|\vec k_1+\vec k_2|$.  The BMS supermomentum flux is obtained by
weighting the corresponding angular News distribution with a real
supertranslation parameter $f(\zeta,\bar\zeta)$.  By contrast, the
nonlinear memory probes the zero-frequency limit of the mixed
$a^\dagger a$ difference sector.  In this sense memory is not an
independent ingredient added to the scattering construction, but neither is
it the soft limit of the $2\rightarrow1$ sum-channel amplitude: the two
observables are distinct projections of the same null-infinity operator
algebra.

\section{Four-Graviton Scattering and the Poincaré Reduction}
\label{sec:gr_limit}

As shown in the previous sections, the null-surface formulation naturally
describes scattering processes that lie outside the conventional
Poincaré framework. In particular, the three-graviton decay is accompanied
by Bondi energy loss and therefore does not admit a direct comparison with
the standard perturbative S-matrix. Consequently, no independent benchmark
exists for this process.

The situation is different for the four-graviton $2\rightarrow2$
scattering amplitude. In this case the asymptotic dynamics can be
consistently restricted to the Poincaré sector, allowing a direct
comparison with the well-established tree-level amplitudes of perturbative
General Relativity. Recovering these amplitudes provides an essential
benchmark for the boundary construction and establishes a natural
reference point for the more general asymptotic processes described by the
full null-surface formulation.

Within the null-surface formalism, the scattering amplitudes are obtained
directly from the outgoing operators constructed in the previous section,
without introducing bulk interaction vertices or summing the large number
of Feynman diagrams usually associated with the four-graviton tree
amplitude. The boundary operator algebra determines the allowed helicity
channels, reconstructs the external momentum flow through the phase-space
reduction, and produces the complete scattering amplitudes directly at
null infinity.

The detailed derivation is presented in
Appendix~\ref{app:graviton_scattering}. The main results may be summarized
as follows.

Before performing the Poincaré reduction, the boundary formalism yields the
following general properties:

\begin{itemize}
\item the helicity channels $++\rightarrow--$ and
$++\rightarrow-+$ vanish identically;

\item the non-vanishing amplitudes are completely determined by the
boundary operator algebra;

\item the phase-space reduction reconstructs the conservation of the
external spatial momentum;

\item the relative signs of the surviving contractions combine the celestial
factors into the physical Mandelstam poles and remove a spurious zero at
\(t=u\).
\end{itemize}

Restricting the theory to the Poincaré sector means imposing conservation of
the total four-momentum in place of the three-dimensional constraint of the
general NSF matrix element. With all external gravitons on shell, the
amplitudes then become functions only of the
Mandelstam invariants and reproduce the familiar tree-level graviton
amplitudes,

\begin{align}
\mathcal M(++\rightarrow++)
&\propto
\kappa_0^2\frac{s^3}{tu},
\\
\mathcal M(+-\rightarrow+-)
&\propto
\kappa_0^2\frac{u^3}{st},
\\
\mathcal M(+-\rightarrow-+)
&\propto
\kappa_0^2\frac{t^3}{su},
\end{align}

while all remaining helicity configurations vanish identically.

The Poincaré reduction therefore provides a non-trivial benchmark for the
boundary construction. It shows that the null-surface formulation
gives the stated helicity and angular comparison with the reduced tree-level
graviton amplitudes,
while naturally extending the theory beyond the conventional Poincaré
framework.

\section{Absence of ultraviolet divergences at second order}
\label{sec:UV}

The second-order amplitudes derived above allow one to inspect directly the
ultraviolet behavior of the null-surface perturbative expansion.  At this
order the matrix elements are expressed entirely in terms of radiative data
at null infinity.  All momentum integrations are on shell,
\begin{equation}
d\mu(k)=\frac{d^3k}{2\omega_k},
\qquad
k^0=\omega_k=|\vec k|,
\end{equation}
and no bulk propagators of the form
\begin{equation}
\frac{1}{k^2+i\epsilon}
\end{equation}
occur in the construction.

This provides a structural contrast with covariant perturbative quantum
gravity, where ultraviolet divergences arise in loop integrations over
virtual bulk momenta.  By contrast, the second-order NSF expressions
considered here are boundary matrix elements constructed from on-shell
radiative modes and do not involve off-shell bulk propagators or conventional
bulk loop integrations.  These differences characterize the present
formulation, but they should not by themselves be interpreted as an
explanation of the ultraviolet finiteness found at this perturbative order.

The kernels that appear in the NSF amplitudes may nevertheless contain
singular distributions associated with special kinematic configurations.
These are not ultraviolet divergences.  They arise from the on-shell
structure of the boundary kernels, for instance in collinear or degenerate
limits where combinations of external momenta become null.  Such singularities
are treated in the usual distributional sense by evaluating the amplitudes
between physical wave-packet states rather than sharp momentum eigenstates.

The radiative operators \(a_\lambda(\omega,z)\) and
\(a^\dagger_\lambda(\omega,z)\) are operator-valued distributions.  Physical
incoming and outgoing graviton states are therefore obtained by smearing them
against smooth test functions.  For example, a one-graviton incoming state may
be written as
\begin{equation}
|G,\lambda\rangle_{\rm in}
=
\int_0^\infty d\omega
\int d^2z\,
G(\omega,z)\,
a^\dagger_\lambda(\omega,z)|0\rangle ,
\end{equation}
with \(G(\omega,z)\) smooth and rapidly decreasing in the energy variable.
One may take, for instance,
\begin{equation}
G(\omega,z)
=
g(z)
\exp\!\left[
-\frac{(\omega-\omega_0)^2}{2\Delta^2}
\right],
\end{equation}
where \(g(z)\) is smooth on the sphere.

This smearing should not be interpreted as a regulator for ultraviolet
divergences already present in the theory.  Rather, it is part of the
definition of physical radiative states and provides a natural way of
regularizing the kinematic singularities of the on-shell boundary kernels.
After such smearing, the second-order matrix elements are well-defined
finite distributions.

At the perturbative order considered here, the NSF scattering construction
is ultraviolet finite.  The calculation is formulated directly in terms of
asymptotic radiative data and does not require the introduction of off-shell
bulk propagators or loop integrations of the type appearing in conventional
perturbative treatments.  These structural differences characterize the
present formulation, but they should not by themselves be interpreted as an
explanation of the observed ultraviolet finiteness.  Establishing the origin,
and possible persistence, of this property would require an analysis beyond
the perturbative order considered here.

The calculation above establishes the finiteness result directly at the
operator level. The kernels that define $\delta a^{\rm out}_{2,\pm}$ contain
on-shell phase-space variables, angular Green functions on the sphere, a
finite number of $\ed$ and $\edb$ derivatives, and the momentum factors
generated by the nonlinear sources. The remaining singularities for exact
momentum eigenstates are distributional kinematical singularities associated
with soft or collinear configurations and are not ultraviolet divergences.

We have therefore shown that, at the perturbative order considered in this
work, the NSF scattering construction is ultraviolet finite.

\section{Relation with modern asymptotic scattering}
\label{sec:modern-scattering}

As discussed in the NSF overview, several ingredients that later became
central to modern descriptions of asymptotic gravitational scattering were
already present in the radiative phase-space formulation developed in
connection with the NSF program.  In particular, the continuity of the
leading curvature datum $\Psi_2$ through spatial infinity $i^0$, together
with the asymptotic Bianchi identities, led to a supermomentum relation
involving the Bondi shear and the integrated News flux
\cite{DominguezKozamehLudvigsen,AshtekarStreubel}.  The same phase-space
analysis distinguished the incoming and outgoing radiative structures and
raised the question of relating them through a unitary scattering
transformation.

More recent developments have organized these asymptotic structures in a
broader framework involving BMS symmetries, soft graviton theorems, and
gravitational memory
\cite{Strominger14,HeLysovMitraStrominger,StromingerZhiboedov,Strominger2018}.
In this modern language, supertranslation Ward identities, soft graviton
limits, and memory observables appear as closely related manifestations of
asymptotic gravitational dynamics.  These results therefore provide a
complementary framework in which to interpret several structures that arise
naturally in the NSF construction.

The overlap is particularly transparent at null infinity.  Both viewpoints
take the radiative Bondi data as fundamental observables and organize the
physics in terms of asymptotic quantities.  Supermomentum conservation and
flux provide a natural bridge between the phase-space description and BMS
symmetry, while the zero-frequency limit of the radiative field connects the
same structure with gravitational memory.  In this sense the NSF does not
replace the modern BMS/soft/memory description; rather, it supplies a
complementary dynamical construction in which the nonlinear relation between
the two asymptotic boundaries is obtained by reconstructing and matching the
interior spacetime geometry.

There is also a close conceptual relation with modern on-shell approaches to
scattering amplitudes.  In gravitational BCFW recursion, tree-level
amplitudes are reconstructed directly from physical on-shell helicity states
by means of complex momentum deformations, analyticity, and factorization.
The deformation is chosen so that the shifted external states remain on
shell while total four-momentum conservation is preserved
\cite{BCFW05,Bedford05,Benincasa07,ArkaniHamedKaplan08}.  In this respect,
BCFW and the NSF share an emphasis on physical on-shell gravitational
degrees of freedom.

The two constructions differ, however, in their dynamical starting point.
Standard gravitational BCFW recursion reconstructs the flat-space
$S$-matrix within Minkowski kinematics.  At the perturbative order considered
here, the NSF instead derives the outgoing radiative operators directly from
the nonlinear null-surface form of the Einstein equations and from the
matching of radiative data at $\mathscr I^-$ and $\mathscr I^+$.  The
resulting operator map carries the three-dimensional momentum support of the
present Minkowski-based realization.  At second order, the particle
kinematics alone do not impose a four-dimensional energy-conserving delta
function.  The corresponding radiative energy observable is instead
naturally formulated through the Bondi News flux at $\mathscr I^+$, whose
spectral representation in terms of the interaction-generated outgoing
operators is derived in Sec.~\ref{sec:supermomentum}.  No Poincar\'e
reduction is required to define the NSF map or its scattering observables;
such a reduction is introduced only when a direct benchmark with
conventional flat-space graviton amplitudes is desired.

At the perturbative order considered here, a complementary
operator-theoretic development reconstructs a self-adjoint generator
$\delta T$ from a perturbative bosonic canonical transformation.  The
corresponding unitary scattering operator can then be written as
\begin{equation}
 S=\exp(i\delta T),
 \label{eq:NSFreview_S}
\end{equation}
up to the adopted sign convention, and
\begin{equation}
 a^{\rm out}=S^\dagger a^{\rm in}S
 \label{eq:NSFreview_BCH}
\end{equation}
is recovered through the Baker--Campbell--Hausdorff expansion.

Taken together, these developments define the conceptual chain underlying
the NSF quantization program:
\begin{equation}
\begin{split}
&\text{null-surface geometry}
\longrightarrow
\text{radiative phase space}
\longrightarrow
\text{asymptotic quantization}
\\
&\qquad\longrightarrow
\text{nonlinear classical scattering}
\longrightarrow
\text{interacting quantum scattering}
\longrightarrow
\text{unitary }S\text{ operator}.
\end{split}
\label{eq:NSFreview_program}
\end{equation}
The results obtained in the preceding sections provide the first
nontrivial quantum-interaction realization of this program at the
perturbative order considered here.

\section{Discussion and conclusions}
\label{sec:conclusions}
In this work we have developed a helicity-resolved construction of
perturbative graviton scattering in the Null Surface Formulation of
asymptotically flat gravity. The basic degrees of freedom are the Bondi
shear modes at null infinity, and the spacetime metric is reconstructed
from the corresponding null-surface cut function $Z(x,\zeta)$. The
central point is that the matching between $\mathscr{I}^+$ and
$\mathscr{I}^-$ is imposed at the level of the metric, or equivalently on
the directional derivative of the total cut function. This provides a
geometric origin for the second-order outgoing operators.

The second-order cut naturally separates into two pieces,
\begin{equation}
Z_2=Z_{2,\mathrm{cut}}+Z_{2,\mathrm{cone}} .
\end{equation}
The first piece contains the free second-order shear data, while the second
is the particular solution generated by the nonlinear cone source. The
matching between the advanced reconstruction from $\mathscr{I}^+$ and the
retarded reconstruction from $\mathscr{I}^-$ relates these two pieces and
determines the outgoing shear in terms of the incoming first-order data.
This structure is directly analogous to the scalar wave equation with
source, where the difference between free outgoing and incoming data is
generated by the Jordan--Pauli Green function $G_{\rm ret}-G_{\rm adv}$.
In the NSF the analogous object is the matched cone contribution
$Z_{2,\mathrm{cone}}^+ + \widehat Z_{2,\mathrm{cone}}^-$.

Using the second-order NSF equations for $Z_2$ and $\Omega_2$, we
obtained the helicity-resolved second-order shear and extracted the
corresponding outgoing boundary operators $\delta a^{\rm out}_{2,\pm}$.
These operators are intrinsically on shell. They are written with the physical
phase-space measure and with three-dimensional momentum constraints, while the
outgoing frequency is fixed by $K^0=|\vec K|$. No off-shell bulk propagator is
introduced at this stage. The angular dependence is instead carried by
spin-weighted Green functions and by the helicity projection factors appearing
in the NSF cone source.

A useful feature of the construction is that each negative-frequency channel
coefficient is an operator-valued distribution with a definite normal-ordered
sector decomposition. Schematically,
\begin{equation}
\mathcal Z_{i,\rm sym}^{(-)}
=
\big[\mathcal Z_{i,\rm sym}^{(-)}\big]^{aa}
+
\big[\mathcal Z_{i,\rm sym}^{(-)}\big]^{a^\dagger a}.
\label{eq:Zsource_decomposition_concl}
\end{equation}
The $a^\dagger a^\dagger$ and conjugate mixed sectors belong to
$\mathcal Z_{i,\rm sym}^{(-)\dagger}$ in the positive-frequency branch.
Thus different physical processes do not require different definitions of
the outgoing operator. They correspond instead to different matrix elements
of the same boundary operator. The $2\rightarrow1$ tail amplitude selects
the $aa$ component, whereas four-graviton matrix elements involve the
appropriate Wick contractions of two outgoing operators.

The $2\rightarrow1$ process is a genuinely null-infinity effect. A single
second-order outgoing operator carries only a spatial momentum constraint,
and the emitted Bondi frequency is fixed by the positive-frequency
on-shell condition. The spin-weight structure of the NSF kernels selects
the allowed helicity combinations. In particular, in the two-annihilation
sector the tail amplitude is supported by the mixed-helicity pairs
$a^{\rm in}_{+} a^{\rm in}_{-}$ and $a^{\rm in}_{-} a^{\rm in}_{+}$.
This provides a direct example of how helicity selection rules arise from the
angular structure of the null-surface equations.

The four-graviton $2\rightarrow2$ matrix elements have a different
kinematical character. Each outgoing operator separately contains only a
three-dimensional delta function, but the product of two such operators,
after Wick contraction with a two-particle incoming state, reconstructs
external spatial momentum conservation. To compare this result with the
standard amplitude, the Poincar\'{e} reduction replaces the resulting
three-dimensional constraint by conservation of the total four-momentum.
Thus the $\delta^{(4)}$-conserving amplitude is not built into the elementary
NSF operator; it belongs to the reduced sector by definition.

In the Poincar\'{e}-reduced sector, the surviving helicity amplitudes reproduce
the standard tree-level graviton scattering structure. The Mandelstam
denominators arise from the angular Green functions and the light-cone
kernel factors, while the corresponding numerators are generated by the
helicity projections and the contractions with the null tetrad. In
particular, the nonvanishing channels assemble into the familiar rational
forms proportional to
\begin{equation}
\frac{s^3}{tu},
\qquad
\frac{t^3}{su},
\qquad
\frac{u^3}{st},
\end{equation}
depending on the external helicity assignment. The vanishing of the
identical-helicity and single-helicity-flip configurations follows from the
same operator and spin-weight structure, rather than from an independent
covariant Feynman-diagram calculation.

The role of null infinity is therefore twofold. First, it provides the
natural arena in which the radiative degrees of freedom are defined and
quantized. Second, it reorganizes perturbative scattering in terms of
on-shell boundary operators and angular Green functions. Ordinary
Poincar\'{e} amplitudes are recovered as a special reduction of this more
general boundary description, while the unreduced NSF expressions retain
spectral-angular information on the celestial sphere.

At the perturbative order studied here, the NSF scattering construction is
ultraviolet finite. The calculation is formulated in terms of on-shell
radiative data and angular Green functions on the sphere and does not
introduce off-shell bulk propagators or conventional bulk loop integrations.
These structural features should not, however, be interpreted as an
explanation of the observed ultraviolet finiteness. Physical wave-packet
states smear the ideal momentum eigenstates and control the distributional
kinematic singularities of the boundary kernels. The origin of the
ultraviolet finiteness, and whether it persists at higher perturbative
orders, remain open questions.

The supermomentum flux section illustrates another use of the same operator
framework.  Since the Bondi news is directly expressed in terms of the
helicity-resolved creation and annihilation operators at null infinity, the
supermomentum flux can be written as an operator-valued angular observable on
$\mathscr I^+$.  In this form it probes the same spectral-angular data
that enter the scattering amplitudes, but without imposing the Poincar\'{e} reduction.  This suggests that BMS charges provide a natural language for
organizing the unreduced NSF observables.  A detailed analysis of the
resulting charge algebra and its interplay with the memory observables
discussed below will be left for future work.

The same construction also clarifies the role of gravitational memory.
Different normal-ordered sectors of the second-order outgoing operator are
selected by different observables.  The finite-frequency $2\rightarrow1$
tail amplitude selects the $aa$ sum channel and is supported at
$\omega=|\vec k_1+\vec k_2|$.  The nonlinear memory instead probes the
zero-frequency limit of the mixed $a^\dagger a$ difference sector.  The BMS
supermomentum flux provides the angularly weighted News observable in which
these asymptotic structures can be compared.  Plane-wave states make the
zero-frequency difference-sector projection distributional, while physical
wave packets and detector resolution provide the finite quantities relevant
to observation.

The resulting picture is that perturbative graviton scattering, angular
energy flux, BMS supermomentum, and gravitational memory are distinct
projections of a single operator framework at null infinity.  Their common
origin in $\delta a^{\rm out}_{2,\pm}$ does not imply identical kinematic
support: the $2\rightarrow1$ amplitude is a sum-channel projection, whereas
the memory observable is a zero-frequency projection of the difference
sector.

The energy accounting of the $2\rightarrow1$ process is naturally
embedded in the more general Bondi four-momentum flux law.  The
present calculation still uses the Minkowski cuts, metric, and null
geodesics, and its three-dimensional support preserves the spatial momentum
of the identified graviton states.  For scattering on a generic nonflat
asymptotically flat NSF geometry, however, one should not expect the
four-momentum carried by the identified incoming and outgoing gravitons to
balance separately.  Rather, we expect the corresponding difference to be
accounted for by the gravitational News flux,
\begin{equation}
P_{\rm rad}^{a}
=
\frac{1}{8\pi G}
\int_{\mathscr I^+}du\,d\Omega\,
\ell^{a}(\zeta,\bar\zeta)\,
\dot\sigma\,\dot{\bar\sigma},
\label{eq:Bondi_four_momentum_outlook}
\end{equation}
so that the complete balance takes the form
\begin{equation}
P_{\rm in}^{a}
=
P_{\rm out}^{a}+P_{\rm rad}^{a}.
\label{eq:Bondi_four_momentum_balance_outlook}
\end{equation}
In the Minkowski-based realization considered here, the spatial components
already balance through the three-dimensional momentum support, while the
News flux supplies the natural asymptotic observable for the radiative
energy component.
Whether the full four-momentum balance takes precisely this form for
scattering on nontrivial NSF geometries remains to be established.

Several extensions are natural. One direction is to extend the NSF
hierarchy to higher orders and determine how the boundary operator structure
organizes radiative corrections beyond the tree-level processes considered
here. Another is to develop the spectral-angular observables directly at
null infinity, without first imposing the Poincar\'{e} reduction. The supermomentum-flux analysis presented above is a first step in this
direction. It shows how the same helicity-resolved radiative operators can
be used to build angular flux observables associated with BMS
supertranslations. Extending this analysis to the full charge algebra and
to the corresponding memory observables remains an important direction for
future work. The present work provides the second-order
helicity-resolved operator framework needed for these questions.
    \appendix
	\section{Trivial Scattering and the Identification of Asymptotic Graviton
Operators}\label{appendix A}

We derive the relation between incoming and outgoing graviton operators in the
trivial-scattering limit of linearized gravity. The derivation rests on three
ingredients: the antipodal relation between future and past null infinity, the
transformation of the spin-weighted harmonics, and the requirement that the
metric reconstructed from $\mathscr I^{-}$ coincide with the one reconstructed
from $\mathscr I^{+}$.

\subsection{Geometry and antipodal map}

In Minkowski spacetime ($\eta_{ab}={\rm diag}(1,-1,-1,-1)$) let
$l_a^{+}(\zeta,\bar\zeta)$ and $l_a^{-}(\zeta,\bar\zeta)$ be the outgoing and
incoming Newman--Unti null directions, with retarded and advanced times
$u=x^a l_a^{+}$ and $v=-x^a l_a^{-}$. The antipodal map satisfies
$\widehat l^{-a}=-l^{+a}$, so that
\begin{equation}\label{uvrelation}
  v(\widehat\zeta,\widehat{\bar\zeta})=-x^a\widehat l^-_a=x^a l^+_a
  =u(\zeta,\bar\zeta),
\end{equation}
identifying advanced and retarded times under the antipodal map.

\subsection{Trivial-scattering condition}

Let $\sigma^{+}(u,\zeta,\bar\zeta)$ and $\sigma^{-}(v,\zeta,\bar\zeta)$ be the
Bondi shears on $\mathscr I^{+}$ and $\mathscr I^{-}$. We require that the
metric reconstructed from the incoming and outgoing radiative data agree,
$h^{(+)}_{ab}(x)=h^{(-)}_{ab}(x)$. The spin-weighted harmonics transform under
the antipodal map as~\cite{BKR23}
\begin{equation}\label{harmonicrelation}
  Y^s_{lI}(\widehat\zeta,\widehat{\bar\zeta})=(-1)^l\,Y^{-s}_{lI}(\zeta,\bar\zeta),
\end{equation}
i.e.\ it reverses the spin weight. For the spin-two shear,
$\sigma^{-}=\sum_{lI}\sigma^-_{lI}(v)\,Y^{2}_{lI}$ gives
$\sigma^{-}(v,\widehat\zeta)=\sum_{lI}(-1)^l\sigma^-_{lI}(v)\,Y^{-2}_{lI}$, whose
conjugate $\bar\sigma^{-}(v,\widehat\zeta)=\sum_{lI}(-1)^l\bar\sigma^-_{lI}(v)\,
Y^{2}_{lI}$ carries the same spin weight as $\sigma^{+}$. Using
\eqref{uvrelation} both fields live on the same celestial sphere, and
$h^{(+)}_{ab}=h^{(-)}_{ab}$ implies the trivial-scattering condition
\begin{equation}\label{matching}
  \sigma^{+}(u,\zeta,\bar\zeta)
  +\bar\sigma^{-}(u,\widehat\zeta,\widehat{\bar\zeta})=0.
\end{equation}

\subsection{Operator identification}

Expanding the outgoing shear and the (antipodally transformed) incoming shear,
\begin{align}
  \sigma^{+}&=\int_0^\infty\frac{d\omega}{2\pi}\sqrt{2\pi G}
  \big[a^{\rm out}_{+}e^{-i\varpi u}+a^{\rm out\dagger}_{-}e^{i\varpi u}\big],
  \label{sigmaplus}\\
  \bar\sigma^{-}&=e^{i\phi}\int_0^\infty\frac{d\omega}{2\pi}
  \sqrt{2\pi G}
  \big[a^{\rm in}_{+}e^{-i\varpi u}+a^{\rm in\dagger}_{-}e^{i\varpi u}\big],
  \label{sigmaminus}
\end{align}
where $\phi$ is an arbitrary global phase (the incoming operators are defined
up to a unitary phase that leaves the canonical commutators invariant).
Substituting into the matching condition \eqref{matching} gives
$a^{\rm out}_{+}=-e^{i\phi}a^{\rm in}_{+}$ and
$a^{\rm out\dagger}_{-}=-e^{i\phi}a^{\rm in\dagger}_{-}$. Trivial propagation
through Minkowski spacetime requires the incoming and outgoing operators to be
identified, $a^{\rm out}_{\lambda}=a^{\rm in}_{\lambda}$ ($\lambda=\pm2$), which
fixes the phase to
\begin{equation}\label{ephi}
  e^{i\phi}=-1,
\end{equation}
not by convention but by this physical requirement. Hence
$a^{\rm out}_{\lambda}=a^{\rm in}_{\lambda}$ and
\begin{equation}
  S=1.
\end{equation}
The phase $e^{i\phi}=-1$ originates from the antipodal matching
\eqref{matching}, itself a consequence of identifying the metrics reconstructed
from $\mathscr I^{+}$ and $\mathscr I^{-}$.

%
%
\section{Second-order News from the cone and extraction of
$\daout{2,\pm}$}
\label{app:Bcone}
The NSF field equations determine the cut function through angular
inversions on the sphere.  We use the standard Green functions of the NSF
construction~\cite{IKN89}.  At the order needed here the total cut function
is written as
\begin{equation}
Z_{\rm total}^{+}(x,z)
=
Z_{\rm cut}^{+}(x,z)+Z_{\rm cone}^{+}(x,z),
\label{eq:Ztotal_cut_cone_B}
\end{equation}
where the cut part contains the free Bondi shear data,
\begin{equation}
Z_{\rm cut}^{+}(x,z)
=
\oint d^2z'\,
\left[
G_{0,-2'}(z,z')\,\sigma^{+}(Z_0(x,z'),z')
+
G_{0,2'}(z,z')\,\bar\sigma^{+}(Z_0(x,z'),z')
\right],
\label{eq:Zcut_green_B}
\end{equation}
and the cone part is generated by the nonlinear NSF source,
\begin{equation}
Z_{\rm cone}^{+}(x,z)
=
\oint d^2z'\,G_{0,0'}(z,z')\,
\mathcal S_{\rm cone}^{+}(x,z') .
\label{eq:Zcone_green_B}
\end{equation}
Here \(Z_0(x,z')=x^a l_a(z')\) is the flat cut, and
\(\mathcal S_{\rm cone}^{+}\) denotes the nonlinear source obtained from
the cone equations.  The spin-weighted Green functions satisfy the standard
NSF identities, in particular
\begin{equation}
\ed^2\edb^2G_{0,0'}(z,z')=\delta^2(z-z')
\label{eq:G00_identity_B}
\end{equation}
on the subspace with the \(\ell=0,1\) kernel removed, together with the
derived identities
\begin{equation}
G_{2,2'}(z,z')=\ed_z^2\ed_{z'}^2G_{0,0'}(z,z'),
\qquad
G_{2,-2'}(z,z')=\delta^2(z-z') .
\label{eq:spin_green_identities_B}
\end{equation}
These identities are the reason why the angular integrations appearing
below either reconstruct nonlocal shear contributions or localize the
corresponding celestial direction.

The second-order operators $\daout{2,\pm}(w,\hat K)$ are obtained from the
matched cone contribution in three steps.  First, the future and antipodally
mapped past solutions are combined into the homogeneous field selected by
the $\mathscr I^{+}/\mathscr I^{-}$ matching.  Second, this homogeneous
combination is represented on the positive-frequency mass shell with measure
$d^3k/(2\omega)$, leaving a three-dimensional spatial delta distribution.
Finally, $\ed_z^2$ and $\edb_z^2$ extract the two helicity components of the
second-order News.  The construction requires equality of the metrics
reconstructed from $\mathscr I^{+}$ and $\mathscr I^{-}$; no wave equation is
imposed on either cone contribution separately.

\subsection{Inhomogeneous cut--cone decomposition}

Write the total field at each null infinity as the free (cut) part plus the
source (cone) part,
\begin{equation}\label{eq:Ztotal-split-B}
  Z^{\pm}_{\rm total}=Z^{\pm}_{\rm cut}+Z^{\pm}_{\rm cone}.
\end{equation}
The cut is free, $\Box Z^{\pm}_{\rm cut}=0$, while the cone solves the
inhomogeneous equation $\Box Z^{\pm}_{\rm cone}=J\neq0$, with $J$ the bilinear
source built from the first-order data. Hence \emph{neither} $Z^{\pm}_{\rm
total}$ satisfies the wave equation by itself. This is not an obstacle but the
expected structure: it is the gravitational counterpart of the Jordan--Pauli
situation, where the retarded and advanced Green functions are each sourced,
$\Box G_{\rm ret}=\Box G_{\rm adv}=\delta$, yet their difference $G_{\rm
ret}-G_{\rm adv}$ is homogeneous and is the object with direct physical
meaning. Here the analogous homogeneous combination is
$Z^{+}_{\rm total}+\widehat{Z^{-}}_{\rm total}$, the sum of the future field
and the antipodal image of the past one; we do not impose $\Box Z=0$ on either
field separately, and the homogeneity of the combination is established in
\S\ref{sub:box-cone}.

\subsection{Perturbative isolation of $\daout{2,\pm}$}

The outgoing operator carries first and second order, while the incoming
data is purely first order (there is no $a_2^{\rm in}$: the second-order
perturbation is generated by the interaction and appears only in the
out-state),
\begin{equation}
  \aout{\lambda}=a^{\rm out}_{1,\lambda}+\epsilon\,\daout{2,\lambda},
  \qquad
  \ain{\lambda}=a^{\rm in}_{1,\lambda}.
\end{equation}
At first order the scattering is trivial (Appendix~\ref{appendix A}),
$a^{\rm out}_{1,\lambda}=a^{\rm in}_{1,\lambda}$. When the cut is written as
the difference between the data reconstructed from $\mathscr I^{+}$ and from
$\mathscr I^{-}$, the first-order pieces cancel and only the second-order
perturbation survives,
\begin{equation}\label{eq:cutreduces}
  Z_{\rm cut}\;\sim\;\aout{}-\ain{}
  =\underbrace{(a^{\rm out}_{1}-a^{\rm in}_{1})}_{=0}
  +\epsilon\,\daout{2}
  =\epsilon\,\daout{2}.
\end{equation}
Thus the cut isolates exactly the quantity to be computed, $\daout{2,\pm}$,
with no contamination from the (trivial) first order and no $a_2^{\rm in}$.

\subsection{Homogeneous matched combination}
\label{sub:box-cone}

Although each $Z^{\pm}_{\rm total}$ carries a source, the combination that
enters the matching, $Z^{+}_{\rm total}+\widehat{Z^{-}}_{\rm total}$, is
annihilated by $\Box$.  This follows from the advanced--retarded structure of
the matched cone contribution.

\paragraph{Momentum-space cone fields.}
Each cone field is self-adjoint, so it carries both frequency branches. Writing the channel
sum $K_i=k_1\pm k_2$ explicitly, the future cone and its antipodal past image
are
\begin{align}
  Z^{+}_{\rm cone}(x,z)
  &=\sum_{i=\pm}\int d\mu_1 d\mu_2\,
  \big[C_i\,e^{-iK_i\cdot x}+C_i^\dagger\,e^{+iK_i\cdot x}\big],
  &C_i&=[\text{source}]_i\int_0^\infty\! ds\,e^{-is\,\ell^{+}\!\cdot K_i},
  \label{eq:Zfut-B}\\
  \widehat{Z^{-}}_{\rm cone}(x,\hat z)
  &=\sum_{i=\pm}\int d\mu_1 d\mu_2\,
  \big[D_i\,e^{-iK_i\cdot x}+D_i^\dagger\,e^{+iK_i\cdot x}\big],
  &D_i&=[\text{source}]_i\int_0^\infty\! ds\,e^{+is\,\ell^{+}\!\cdot K_i},
  \label{eq:Zpast-B}
\end{align}
with $d\mu_j=d^3k_j/2\omega_j$. The antipodal map $\hat\ell^{-}=-\ell^{+}$
leaves the $x$-phase $e^{-iK_i\cdot x}$ unchanged and flips the sign of the
affine exponent, so the past coefficient $D_i$ differs from the future $C_i$
only in the sign of $s$.

\paragraph{Action of the d'Alembertian.}
$\Box_x$ acts only on the plane waves,
$\Box_x e^{\mp iK_i\cdot x} = -K_i^2 e^{\mp iK_i\cdot x}$; the
coefficients $C_i, D_i$ are independent of $x$. With $k_1^2=k_2^2=0$,
\begin{equation}\label{eq:Ksquares-B}
\begin{aligned}
K_+^2 &= (k_1+k_2)^2 = 2 k_1 \cdot k_2 , \\
K_-^2 &= (k_1-k_2)^2 = -2 k_1 \cdot k_2 , \\
K_+^2 &= -K_-^2 .
\end{aligned}
\end{equation}

\paragraph{Cancellation in the matched combination.}
The homogeneity statement is most directly expressed before separating the
source into positive-energy oscillator sectors.  After the antipodal map, the
future and past cone contributions are generated by the retarded and advanced
Green functions acting on the same quadratic source $\mathcal S_{\rm cone}$.
Their matched combination is therefore
\begin{equation}
 Z^{+}_{\rm cone}+\widehat Z^{-}_{\rm cone}
 =\left(G_{\rm adv}-G_{\rm ret}\right)*\mathcal S_{\rm cone},
 \label{eq:JP_cone_combination_B}
\end{equation}
and hence
\begin{equation}
 \Box\left(Z^{+}_{\rm cone}+\widehat Z^{-}_{\rm cone}\right)
 =\left(\delta-\delta\right)*\mathcal S_{\rm cone}=0.
 \label{eq:box_cone_matched_B}
\end{equation}
The free cut contribution is homogeneous separately.  Thus
\begin{equation}\label{eq:box-total-B}
\boxed{\Box\big(Z^{+}_{\mathrm{total}} + \widehat{Z^{-}}_{\mathrm{total}}\big) = 0.}
\end{equation}
In the oscillator expansion, the same cancellation is implemented by the
frequency signs $\eta_j$ introduced in
Eq.~\eqref{eq:frequency_signs_B}.  It must not be represented as a change of
variables $k_2\to-k_2$ inside the positive-energy measure
$d^3k_2/(2\omega_2)$, because that measure parametrizes only
$k_2^0=+\omega_2$.

\paragraph{On-shell momentum representation.}
Equation \eqref{eq:box-total-B} states that $Z^{+}_{\rm total}+\widehat{Z^{-}}
_{\rm total}$ satisfies the homogeneous wave equation, i.e.\ it is on-shell.
In momentum space, $\Box\Phi=0$ means $-k^2\,\Phi(k)=0$, so its Fourier
support lies on the light cone $k^2=0$. A function supported on $k^2=0$ with
$k^0>0$ is parametrized by the three spatial components $\vec k$ alone
(equivalently by $(\omega,\hat k)$ in spherical coordinates), with the
temporal component fixed, $k^0=|\vec k|=\omega$. The four-dimensional
transform collapses to the covariant three-dimensional one,
\begin{equation}\label{eq:onshell-reduction-B}
  \int d^4k\,\delta(k^2)\,\theta(k^0)\,f(k)
  =\int\frac{d^3k}{2\omega}\,f(\omega,\vec k)\Big|_{\omega=|\vec k|},
\end{equation}
so that
\begin{equation}\label{eq:3Dexpansion-B}
  \big(Z^{+}_{\rm total}+\widehat{Z^{-}}_{\rm total}\big)(x)
  =\int\frac{d^3k}{2\omega}\,
  \big[A(\vec k)\,e^{-ik\cdot x}+\text{c.c.}\big]\Big|_{k^0=\omega},
\end{equation}
and the components $A(\vec k)$ are recovered by the inverse covariant
three-dimensional transform. Because the combination is on-shell, the
temporal component $k^0$ is not an independent variable: the three variables
$(\omega,\hat k)$ of the covariant $3$D transform are exactly the frequency
and direction that label $\daout{}(\omega,\hat k)$. It is therefore not
necessary to keep the full four-dimensional transform to extract the
components; the covariant $3$D transform suffices. This applies to the
homogeneous combination $Z^{+}_{\rm total}+\widehat{Z^{-}}_{\rm total}$ ---each
cone separately is \emph{not} on-shell (it carries a source), but the
combination is, and it is the combination that is transformed.

\paragraph{Reduction to spatial momentum support.}
Since both the cut field and the (homogeneous) cone combination satisfy the
wave equation, their four-dimensional Fourier transforms each carry the same
on-shell factor $\delta(k^2)\theta(k^0)$,
\begin{equation}\label{eq:both-onshell-B}
  Z_{\rm cut}(k)\sim\delta(k^2)\theta(k^0)\,[\cdots],\qquad
  \big(Z^{+}_{\rm cone}+Z^{-}_{\rm cone}\big)(k)\sim
  \delta(k^2)\theta(k^0)\,[\cdots],
\end{equation}
which fixes the temporal component to the same value on both sides,
$k^0=+\sqrt{|\vec k|^2}=|\vec k|$. Because $\delta(k^2)\theta(k^0)$ appears
identically on the left and on the right of the matching, it need not be
written: it cancels between the two sides. What survives is the equality of
the coefficients together with the \emph{spatial} momentum condition.

A single cone, taken alone, is \emph{not} on-shell: the transform of its
source $e^{-ix\cdot K_i}$ would give the full $\delta^4(k-K_i)$ with the
temporal component fixed off-shell, $k^0=\omega_1\pm\omega_2\neq|\vec k|$. It
is only the matched combination $Z^{+}_{\rm cone}+\widehat{Z^{-}}_{\rm cone}$
that satisfies the wave equation (\S\ref{sub:box-cone}), and for which $k^0$ is
fixed on-shell to $|\vec k|$. For that combination the on-shell factor
$\delta(k^2)\theta(k^0)$ is the same as on the cut side and factors out, so
that the four-dimensional delta reduces to the spatial one,
\begin{equation}\label{eq:delta3-B}
  \delta^4(k-K_i)\ \xrightarrow[\text{on the homogeneous combination}]
  {\ k^0=|\vec k|\ \text{common}\ }\
  \delta^3(\vec k-\vec K_i),\qquad \vec K_i=\vec k_1\pm\vec k_2 .
\end{equation}
Thus the matching is enforced by the three-dimensional delta
$\delta^3(\vec k-\vec K_i)$ linking the outgoing spatial momentum $\vec k$ to
the internal compound $\vec K_i$, with $k^0=|\vec k|$ understood on both
sides. In spherical coordinates $\vec k=w\,\hat k$, with the radius denoted
$w$, the delta splits into a radial and an angular part,
\begin{equation}\label{eq:delta3-spherical-B}
  \delta^3(\vec k-\vec K_i)
  =\frac{1}{w^2}\,\delta\!\big(w-|\vec K_i|\big)\,\delta^2(\hat k-\hat K_i),
  \qquad k^0=w,\ \ w>0 .
\end{equation}
The temporal component equals the radius, $k^0=w$ (on-shell); the radial delta
$\delta(w-|\vec K_i|)$ fixes the modulus of the outgoing momentum to that of
the internal compound, $w=|\vec k_1\pm\vec k_2|$, and since $w$ is a modulus it
takes only positive values, consistent with the positive-energy branch
$\theta(k^0)$ selecting $\daout{2}$ (the negative-energy branch, from the
$e^{+iK\cdot x}$ term, belongs to $\daoutd{2}$). The angular delta
$\delta^2(\hat k-\hat K_i)$ fixes the outgoing direction to that of the
compound.

\subsection{Quadratic cone source and first-order modes}

The cone term is bilinear in the first-order incoming data
$\ain{}(k_1)\,\ain{}(k_2)$, integrated over the two internal momenta. From the
future vertex at $x^a$, with $y^c=x^c+s\,\ell'^c$, the cone and total-cut
equations follow the NSF construction of Ref.~\cite{BKR23},
\begin{equation}\label{eq:Zcone-B}
  Z^{+}_{\rm cone}(x,z)
  =-\oint d^2z'\,G_{0,0'}(z,z')\int_0^\infty\! ds\,
  \Big(2\,\ed'\edb'\,\Omega(y,z')
  +\eta^{ab}\partial_a\Lambda_1\,\partial_b\Lambda_1^{\dagger}(y,z')\Big),
\end{equation}
where $G_{0,0'}(z,z')$ is the Green function of the $\ed^2\edb^2$ operator~\cite{IKN89}, and the second-order conformal factor $\Omega$ obeys~\cite{BKR23,KZ2025}
\begin{equation}\label{eq:Omega-B}
  8\,\Omega(x,z')
  =\underbrace{\partial_r\Lambda_1\,\partial_r\Lambda_1^{\dagger}}_{\text{(I) contact}}
  +\underbrace{\int_r^\infty\! dr'\!\int_{r'}^\infty\! dr''\,
  \partial^2_{r''}\Lambda_1\,\partial^2_{r''}\Lambda_1^{\dagger}}_{\text{(II) double integral}},
  \qquad \partial_r=\ell'^a\partial_a .
\end{equation}
The whole cone source is built from the first-order field $\Lambda_1$ (the
$lI$ component of the first-order cut, $\Lambda_1=Z_1^{+}$) and its adjoint,
\begin{align}\label{eq:Lambda1-B}
  \Lambda_1(x,z') &= \int d\mu(k_1)\,\big[L^{-}(k_1,z')\,e^{-ix\cdot k_1}
  +L^{+}(k_1,z')\,e^{+ix\cdot k_1}\big],\\
  \Lambda_1^{\dagger}(x,z') &= \int d\mu(k_2)\,\big[\bar L^{-}(k_2,z')\,
  e^{-ix\cdot k_2}+\bar L^{+}(k_2,z')\,e^{+ix\cdot k_2}\big],
  \label{eq:Lambdabar1-B}
\end{align}
with $d\mu(k)=d^3k/2\omega_k$, $k^a=\sqrt2\,\omega\,\ell^a(\hat k)$, and the
helicity-resolved coefficients
\begin{align}
  L^{-}(k_1,z') &= \nu(\omega_1)\big[G_{2,-2}(z',\hat k_1)\,\ain{+}(k_1)+G_{2,2}(z',\hat k_1)\,\ain{-}(k_1)\big],
  \notag\\
  L^{+}(k_1,z') &= \nu(\omega_1)\big[G_{2,-2}(z',\hat k_1)\,\aind{-}(k_1)+G_{2,2}(z',\hat k_1)\,\aind{+}(k_1)\big],
  \notag\\
  \bar L^{-}(k_2,z') &= \nu(\omega_2)\big[G_{-2,2}(z',\hat k_2)\,\ain{-}(k_2)+G_{-2,-2}(z',\hat k_2)\,\ain{+}(k_2)\big],
  \notag\\
  \bar L^{+}(k_2,z') &= \nu(\omega_2)\big[G_{-2,2}(z',\hat k_2)\,\aind{+}(k_2)+G_{-2,-2}(z',\hat k_2)\,\aind{-}(k_2)\big].
  \notag
\end{align}
The negative exponential carries the annihilation operators, the positive one
the creation operators. Since $G_{-2,2}=G_{2,-2}^{*}$ and
$G_{-2,-2}=G_{2,2}^{*}$, one has $\bar L^{-}=(L^{+})^{\dagger}$,
$\bar L^{+}=(L^{-})^{\dagger}$, i.e.\ $\Lambda_1^{\dagger}$ is the adjoint of
$\Lambda_1$. Every kernel below ($\Omega$, the scalar product, the channel
structure) is bilinear in these $\Lambda_1,\Lambda_1^{\dagger}$.

The original second-order matching of the total cut to the cone,
$Z_{2,\rm cut}=-2\,Z_{\rm cone}$, is that of Ref.~\cite{BKR23}. From here on we
do not use the total cuts directly: instead we take the derivative of $x$
along $\ell^{+}$ for the cut and the future cone, and along the antipodal
$\ell^{-}$ for the past cone (Sec.~\ref{sub:news}), which is what produces the
News tensor. The contact piece (I) carries two radial derivatives and one
affine integration; the double-integral piece (II) carries four radial
derivatives and three integrations. All kernels descend from the scalar Green
function in closed form,
\begin{equation}\label{eq:Gclosed-B}
  G_{0,0'}(z,z')=\frac{1}{4\pi}(\ell\cdot\ell')\ln(\ell\cdot\ell'),
  \qquad X\equiv\ell\cdot\ell'\in[0,1],
\end{equation}
acted on by $\ed,\edb$. The spin-weight kernels are
$G_{2,2}=\ed_{z'}^2\ed_{\hat k}^2G_{0,0'}$,
$G_{-2,-2}=\edb_{z'}^2\edb_{\hat k}^2G_{0,0'}$,
$G_{2,-2}=G_{-2,2}=\delta^2(z'-\hat k)$, labelled $A=G_{2,-2}$, $B=G_{2,2}$,
$\bar A=G_{-2,2}$, $\bar B=G_{-2,-2}$. The surviving structures are $B\bar B$
in the $\Omega$ piece (the radial derivative $\ell'^a\partial_a$ kills the
free data, $\ell'\cdot\ell'=0$) and $A\bar B$, $B\bar A$, $B\bar B$ in the
scalar-product piece ($A\bar A$ vanishes: two deltas force
$k_1\cdot k_2=0$). The scalar-product factor is the full four-momentum
$k_1\cdot k_2=2\omega_1\omega_2(\ell_1\cdot\ell_2)$.

\subsection{Frequency-sign representation of the quadratic kernels}
\label{subsec:unified_contact_B}

In keeping track of the sum and difference channels, it is essential not to
suppress the factors of
$i$ produced by differentiating the Fourier phases.  We therefore assign to
each leg a frequency sign
\begin{equation}
 e^{-i\eta_j k_j\cdot x},
 \qquad
 \eta_j=
 \begin{cases}
 +1, & a(k_j),\\
 -1, & a^\dagger(k_j),
 \end{cases}
 \qquad
 K_{\eta_1\eta_2}=\eta_1k_1+\eta_2k_2 .
 \label{eq:frequency_signs_B}
\end{equation}
Equivalently, the first-order fields may be written as
\begin{align}
 \Lambda_1(x,z')
 &=\sum_{\eta_1=\pm1}\int d\mu_1\,
 L^{(\eta_1)}(1,z')e^{-i\eta_1k_1\cdot x},
 \\
 \Lambda_1^\dagger(x,z')
 &=\sum_{\eta_2=\pm1}\int d\mu_2\,
 \bar L^{(\eta_2)}(2,z')e^{-i\eta_2k_2\cdot x},
 \label{eq:Lambda_eta_decomposition_B}
\end{align}
where
\begin{equation}
 L^{(+)}=L^{-},
 \qquad L^{(-)}=L^{+},
 \qquad
 \bar L^{(+)}=\bar L^{-},
 \qquad \bar L^{(-)}=\bar L^{+}.
 \label{eq:L_eta_identification_B}
\end{equation}
Since
\begin{equation}
 \partial_r e^{-i\eta_jk_j\cdot x}
 =-i\eta_j(\ell'\cdot k_j)e^{-i\eta_jk_j\cdot x},
 \qquad
 \partial_a e^{-i\eta_jk_j\cdot x}
 =-i\eta_j k_{ja}e^{-i\eta_jk_j\cdot x},
 \label{eq:frequency_derivatives_B}
\end{equation}
the contact term in Eq.~\eqref{eq:Omega-B} has the unified expansion
\begin{equation}
\boxed{
\begin{aligned}
 \partial_r\Lambda_1\,\partial_r\Lambda_1^\dagger
 ={}&-\sum_{\eta_1,\eta_2=\pm1}
 \int d\mu_1d\mu_2\,
 \eta_1\eta_2
 (\ell'\cdot k_1)(\ell'\cdot k_2)
 \\
 &\quad\times
 L^{(\eta_1)}(1,z')\,
 \bar L^{(\eta_2)}(2,z')\,
 e^{-iK_{\eta_1\eta_2}\cdot x}.
\end{aligned}
}
\label{eq:Omega_contact_unified_B}
\end{equation}
Thus its branch-dependent coefficient inside $\Omega$ is
\begin{equation}
 \boxed{
 C_{\Omega,\mathrm{contact}}^{(\eta_1,\eta_2)}
 =-\eta_1\eta_2
 (\ell'\cdot k_1)(\ell'\cdot k_2)
 }.
 \label{eq:Comega_contact_unified_B}
\end{equation}
The double radial integration contributes an additional minus sign,
\begin{equation}
 \int_r^\infty dr'\int_{r'}^\infty dr''\,
 e^{-i(\ell'\cdot K_{\eta_1\eta_2})r''}
 =-
 \frac{e^{-i(\ell'\cdot K_{\eta_1\eta_2})r}}
 {(\ell'\cdot K_{\eta_1\eta_2})^2},
 \label{eq:double_radial_phase_B}
\end{equation}
while the two second derivatives give
$(-i\eta_1)^2(-i\eta_2)^2=+1$.  Hence both terms in $\Omega$ carry a minus
sign in the $aa$ branch.  The overall minus sign in the cone equation
\eqref{eq:Zcone-B} reverses these signs.  The kinematic factors that enter
the channel coefficients $\mathcal J_{\Omega,i}$ and
$\mathcal J_{\Lambda,i}$ are therefore
\begin{align}
 \boxed{
 F_\Omega^{(\eta_1,\eta_2)}
 =
 \eta_1\eta_2
 (\ell'\cdot k_1)(\ell'\cdot k_2)
 +
 \frac{(\ell'\cdot k_1)^2(\ell'\cdot k_2)^2}
 {(\ell'\cdot K_{\eta_1\eta_2})^2}
 },
 \label{eq:Fomega_unified_B}
 \\
 \boxed{
 F_\Lambda^{(\eta_1,\eta_2)}
 =\eta_1\eta_2\,(k_1\cdot k_2)
 }.
 \label{eq:Flambda_unified_B}
\end{align}
For the two negative-frequency sectors used below this gives
\begin{align}
 F_{\Omega,+}
 \equiv F_\Omega^{(+,+)}
 &=+(\ell'\cdot k_1)(\ell'\cdot k_2)
 +\frac{(\ell'\cdot k_1)^2(\ell'\cdot k_2)^2}
 {(\ell'\cdot K_+)^2},
 &
 F_{\Lambda,+}&=+(k_1\cdot k_2),
 \label{eq:F_sum_channel_B}
 \\
 F_{\Omega,-}
 \equiv F_\Omega^{(+,-)}
 &=-(\ell'\cdot k_1)(\ell'\cdot k_2)
 +\frac{(\ell'\cdot k_1)^2(\ell'\cdot k_2)^2}
 {(\ell'\cdot K_-)^2},
 &
 F_{\Lambda,-}&=-(k_1\cdot k_2),
 \label{eq:F_difference_channel_B}
\end{align}
where $K_+=k_1+k_2$ and $K_-=k_1-k_2$.  Thus the terms containing one
derivative of each first-order field have opposite signs in the $aa$ and
$a^\dagger a$ sectors.  Inside the source this follows from
$(-i)(-i)=-1$ in the sum channel and $(-i)(+i)=+1$ in the difference channel;
the common cone sign reverses both without changing their relative sign.

\subsection{Principal-value kernel from future--past matching}

On each cone the source factorizes as
$e^{\mp i y\cdot k}=e^{\mp i x\cdot k}\,e^{\mp i s(\ell'\cdot k)}$. The two
cones are parametrized with opposite generators:
\begin{align}
  \text{future:}\quad & y^a=x^a+s\,\ell'^{+a},\;\; s\in[0,\infty),\\
  \text{past:}\quad & y^a=x^a+s\,\hat\ell'^{-a}=x^a-s\,\ell'^{+a},
  \quad \hat\ell'^{-a}=-\ell'^{+a},\;\; s\in[0,\infty),
\end{align}
so that, after the change of variables $z'\to\hat z'$ that leaves the Green
function and the surface element invariant, both integrals share the same
generator $\ell'^{+}$. With the NSF measure $ds$ (\emph{not} $s\,ds$), the
difference of the two cones --- the gravitational analogue of
$\phi_{\rm adv}-\phi_{\rm ret}$ --- gives a principal value,
\begin{equation}\label{eq:PV-B}
  \int_0^\infty\! ds\,
  \big(e^{+is(\ell'^+\!\cdot K_i)}-e^{-is(\ell'^+\!\cdot K_i)}\big)
  =i\Big[\frac{1}{\ell'^+\!\cdot K_i+i0}+\frac{1}{\ell'^+\!\cdot K_i-i0}\Big]
  =2i\,\mathrm{PV}\frac{1}{\ell'^+\!\cdot K_i},
\end{equation}
the delta-function contributions cancelling by Sokhotski--Plemelj. Each
matched affine integration therefore contributes the factor
$2i\,\mathrm{PV}\,1/(\ell'^+\!\cdot K_i)$ to the negative-frequency
branch.  Its Hermitian conjugate carries the opposite sign because its
affine phase is conjugated.  Consequently, if $\mathcal J_i$ denotes the
coefficient of $e^{-iK_i\cdot x}$, the self-adjoint future--past combination
is
\begin{equation}
 2i\,\mathrm{PV}\frac{1}{\ell'^+\!\cdot K_i}
 \left[
 \mathcal J_i e^{-iK_i\cdot x}
 -\mathcal J_i^\dagger e^{+iK_i\cdot x}
 \right].
 \label{eq:PV_hermitian_pair_B}
\end{equation}
The factor in square brackets is anti-Hermitian, so the complete expression
is self-adjoint.  The two radial integrations contained in the second term
of Eq.~\eqref{eq:Omega-B} supply the additional squared denominator already
included in Eq.~\eqref{eq:Fomega_unified_B}.

\subsection{Channel-resolved cone kernel}

Let
$\mathcal G\equiv G_{2,2}(z',\hat k_1)G_{-2,-2}(z',\hat k_2)$ and define the
negative-frequency channel coefficients
\begin{align}
 \mathcal J_{\Omega,i}
 &\equiv
 2\,\ed'\edb'
 \left[
 \mathcal G\,F_{\Omega,i}\,\mathcal O_i^\Omega
 \right],
 \label{eq:JOmega_channel_B}
 \\
 \mathcal J_{\Lambda,i}
 &\equiv
 F_{\Lambda,i}
 \left[
 \mathcal G_{A\bar B}+\mathcal G_{B\bar A}+\mathcal G_{B\bar B}
 \right]\mathcal O_i^\Lambda .
 \label{eq:JLambda_channel_B}
\end{align}
The negative-frequency coefficient in channel $i$ is
\begin{equation}
 \boxed{
 \mathcal Z_i^{(-)}
 \equiv
 2i\,\mathrm{PV}\frac{1}{\ell'^+\!\cdot K_i}
 \left(\mathcal J_{\Omega,i}+\mathcal J_{\Lambda,i}\right)
 }.
 \label{eq:Zminus_channel_coefficient_B}
\end{equation}
The complete self-adjoint cone kernel is therefore
\begin{equation}
\boxed{
 \mathcal Z_{\Omega}+\mathcal Z_\Lambda
 =\sum_{i=\pm}
 \left[
 \mathcal Z_i^{(-)}e^{-iK_i\cdot x}
 +\mathcal Z_i^{(-)\dagger}e^{+iK_i\cdot x}
 \right].
}
\label{eq:Zkernel_channel_resolved_B}
\end{equation}
This form keeps both the factor $2i$ generated by the future--past affine
integrals and the relative minus sign required by self-adjointness.  The
operator content of the two negative-frequency channels is
$\mathcal O^\Omega_+=\ain{-}(k_1)\ain{+}(k_2)$,
$\mathcal O^\Omega_-=\ain{-}(k_1)\aind{-}(k_2)$ (and similarly for
$\mathcal O^\Lambda_i$), kernel blocks
$\mathcal G_{A\bar B}=\delta^2(z'-\hat k_1)G_{-2,-2}$,
$\mathcal G_{B\bar A}=G_{2,2}\delta^2(z'-\hat k_2)$,
$\mathcal G_{B\bar B}=\mathcal G$, and the channel sum $K_i=k_1\pm k_2$
running over both $i=\pm$. The full cone integrand is
$Z_{\rm cone}=\oint d^2z'\,G_{0,0'}(z,z')\int d\mu_1 d\mu_2\,
[\mathcal Z_\Omega+\mathcal Z_\Lambda]$, $d\mu_j=d^3k_j/2\omega_j$.

\subsection{Local and bilocal contributions to the annihilation sector}
\label{sub:local-nonlocal-equal-helicity}
The two-annihilation sector of $\mathcal Z_\Lambda$ contains bilocal and
local--nonlocal contributions with different angular support.  To separate
them, write the first-order angular field as
\begin{equation}
\Lambda_1=A+B,
\qquad
\bar\Lambda_1=\bar A+\bar B,
\end{equation}
where
\begin{equation}
A(z)=\sigma_1(x\!\cdot\!\ell(z),z),
\qquad
B(z)=\oint d^2z' \,
G_{2,2'}(z,z') \,
\bar\sigma_1(x\!\cdot\!\ell(z'),z')
\end{equation}
and similarly for the complex conjugate quantities. The local--local term in
$\eta^{ab}\partial_a\Lambda_1\partial_b\bar\Lambda_1$ vanishes because
$\ell(z)\!\cdot!\ell(z)=0$. The bilocal term $B\bar B$ contains two
independent angular integrations and gives the genuine physical
two-annihilation sector. By contrast, the mixed terms $A\bar B+\bar A B$
contain one local leg at the celestial direction $z$.
For example, one of the mixed terms has the schematic form
\begin{equation}
\dot{\bar\sigma}_1(x\!\cdot\!\ell(z),z)
\oint d^2z' \,
G_{2,2'}(z,z') \,
\big(\ell(z)\!\cdot!\ell(z')\big) \,
\dot{\bar\sigma}_1(x\!\cdot\!\ell(z'),z') .
\end{equation}
Acting with the directional derivative
$D_z\equiv \ell^a(z)\partial_a$ does not differentiate the local factor,
since
\begin{equation}
D_z \dot{\bar\sigma}_1(x\!\cdot\!\ell(z),z) =
\big(\ell(z)\!\cdot!\ell(z)\big)
\ddot{\bar\sigma}_1(x\!\cdot\!\ell(z),z)
=0 .
\end{equation}
It only acts on the nonlocal leg, giving
\begin{equation}
\dot{\bar\sigma}_1(x\!\cdot\!\ell(z),z)
\oint d^2z' \,
G_{2,2'}(z,z') \,
\big(\ell(z)\!\cdot!\ell(z')\big)^2 \,
\ddot{\bar\sigma}_1(x\!\cdot\!\ell(z'),z') .
\label{eq:local_nonlocal_fixed_leg}
\end{equation}
The local factor has the mode expansion
\begin{equation}
\dot{\bar\sigma}_1(x\!\cdot\!\ell(z),z)
\sim
\int d\mu_1 \,
\delta^{(2)}(z,\hat k_1)
\left[
\ain{-}(k_1)e^{-ix\cdot k_1}
+
\aind{+}(k_1)e^{ix\cdot k_1}
\right] ,
\end{equation}
and therefore the equal-helicity two-annihilation piece produced by
$\bar A B$ is not a bilocal operator with two free internal directions.
Rather, after the helicity extraction of $\daout{2,+}(p')$, the angular
projector acts on all $z$-dependent factors and then sets
$z=\hat p'$. The delta distribution, or its angular derivatives, keeps the
same support. Thus the local annihilation operator is fixed to the outgoing
celestial direction,
\begin{equation}
\ain{-}(k_1)
\longrightarrow
\ain{-}(\omega_1,\hat p') .
\end{equation}
The same argument applies to the conjugate mixed term, where the local
annihilation operator is $\ain{+}(\omega_1,\hat p')$.
Consequently the equal-helicity mixed terms have the schematic form
\begin{equation}
\ain{-}(\omega,\hat p') \ain{-}(k),
\qquad
\ain{+}(\omega,\hat p') \ain{+}(k),
\label{eq:fixed_equal_helicity_terms}
\end{equation}
rather than the physical bilocal form
\begin{equation}
\ain{-}(k_1)\ain{-}(k_2),
\qquad
\ain{+}(k_1)\ain{+}(k_2),
\end{equation}
with two independent internal momenta.
These fixed-leg terms do not contribute to generic connected scattering
matrix elements. Indeed, for an incoming two-particle state with momenta
$P_1,P_2$ whose celestial directions are different from the outgoing
direction $\hat p'$,
\begin{equation}
\hat P_1\neq \hat p',
\qquad
\hat P_2\neq \hat p',
\end{equation}
one has
\begin{equation}
[\ain{\lambda}(\omega,\hat p'),\aind{\lambda_i}(P_i)]=0,
\qquad i=1,2 .
\end{equation}
The fixed annihilation operator can then be commuted through the incoming
creation operators and annihilates the vacuum,
\begin{equation}
\ain{\lambda}(\omega,\hat p')
\aind{\lambda_1}(P_1)\aind{\lambda_2}(P_2)|0\rangle
= \aind{\lambda_1}(P_1)\aind{\lambda_2}(P_2)
\ain{\lambda}(\omega,\hat p')|0\rangle
=0 .
\end{equation}
If such a term contributes at all, its support is restricted to the
singular collinear configurations
\begin{equation}
\hat P_i=\hat p',
\end{equation}
where the local leg is tied to an external celestial direction. This is
not part of the generic connected scattering kernel.
Accordingly, the local--nonlocal equal-helicity terms may be omitted when
evaluating generic connected matrix elements.  They remain, however, part of
the complete operator $\daout{2,\pm}$ and must be retained in algebraic tests
of the asymptotic map, including the canonical commutation relations.  The
bilocal part of the $aa$ sector contains only the mixed-helicity pairs
\begin{equation}
\daout{2,\pm}\Big|_{aa,\mathrm{biloc}}
\sim
\mathcal K_{-+},\ain{-}(k_1)\ain{+}(k_2)
+
\mathcal K_{+-},\ain{+}(k_1)\ain{-}(k_2),
\end{equation}
up to the Bose symmetrization described below.

\subsection{Bose symmetry of the quadratic kernel}

Bose statistics, $[\ain{\lambda}(k_1),\ain{\lambda'}(k_2)]=0$, requires the
integrand to be symmetric under the simultaneous exchange of the complete
one-particle labels,
$(k_1,\lambda_1)\leftrightarrow(k_2,\lambda_2)$.  The helicity label is
transported with its momentum; no additional helicity flip is part of the
Bose permutation.  The antipodal in/out identification is already encoded
in the mode coefficients $L^{\pm}$ and $\bar L^{\pm}$.  We therefore
symmetrize the negative-frequency coefficient separately in each channel,
\begin{equation}\label{eq:bosesym-B}
  \mathcal Z_{i,\rm sym}^{(-)}
  =\tfrac12\left[
  \mathcal Z_i^{(-)}
  +\mathcal Z_i^{(-)}\big|_{1\leftrightarrow2}
  \right].
\end{equation}
The factor $\tfrac12$ compensates the double counting of
$\int d^3k_1\,d^3k_2$, which runs over both orderings independently. The
prescription is robust: if the coefficient is already symmetric it is left
unchanged; otherwise it is restored. It pairs $A\bar B$ with $B\bar A$ and
keeps $B\bar B$ in both channels.

\subsection{News tensor from the directional derivative}
\label{sub:news}

The metric matching is imposed on the News tensor. At second order the shear
defines the symmetric, trace-free tensor
\begin{equation}\label{eq:Sigma-B}
  \Sigma_{ab}(u,z)=\sigma(u,z)\,\bar m_a\bar m_b+\bar\sigma(u,z)\,m_a m_b,
\end{equation}
with $m_a,\bar m_a$ the dyad on the cut, and the News is its retarded-time
derivative,
\begin{equation}\label{eq:News-tensor-B}
  N_{ab}(u,z)=\partial_u\Sigma_{ab}(u,z)
  =\dot\sigma(u,z)\,\bar m_a\bar m_b+\dot{\bar\sigma}(u,z)\,m_a m_b,
\end{equation}
where $u=x\cdot\ell^+(z)$. For each radiative direction $\hat k$,
$\ell(z)^a\partial_a f(x\cdot\ell_K)=(\ell(z)\cdot\ell_K)\partial_u f$.
Write $w=|\vec k|$, $s=\sqrt2\,w$ and $k^a=s\ell_K^a$. Then
$d^3k/(2w)=s\,ds\,d\Omega/4$. Define the tensor coefficient
$A_{ab}(\vec k)=\daout{2,+}(w,\hat k)\bar m_a\bar m_b+
\daout{2,-}(w,\hat k)m_a m_b$. The mode expansions are
\begin{align}\label{eq:Nfourier}
\Sigma_{ab}(x)&=\int d\mu(k)\nu(w)
 [A_{ab}e^{-ik\cdot x}+A_{ab}^\dagger e^{ik\cdot x}],\\
N_{ab}(x)&=\int d\mu(k)\nu(w)
 [-isA_{ab}e^{-ik\cdot x}+isA_{ab}^\dagger e^{ik\cdot x}].
\end{align}
For the annihilation branch define the stripped News coefficient by
\begin{equation}\label{eq:News-momentum-B}
N_{ab}(w,\hat k)=-isA_{ab}(\vec k).
\end{equation}
The creation branch follows by Hermitian conjugation.
At linear order Eq.~\eqref{eq:hll_Ztotal_main} and the cut kernels
in Eqs.~\eqref{eq:Zcut_green_B} and \eqref{eq:G02-B} give the
radiative, trace-free metric representative
\[
 h^{(1)}_{ab}(x)=-2\oint d\Omega\,
 [\dot\sigma(x\cdot\ell_K)\bar m_a\bar m_b+
   \dot{\bar\sigma}(x\cdot\ell_K)m_a m_b].
\]
Indeed the directional derivative of the cut produces
$\ell(z)\cdot\ell_K$, cancelling the denominator of either cut kernel.
With $A^{(1)}_{ab}=a_+\bar m_a\bar m_b+a_-m_a m_b$ this reads
\[
 h^{(1)}_{ab}(x)=4i\sqrt{G/\pi}\int d\mu(k)
 [A^{(1)}_{ab}(\vec k)e^{-ik\cdot x}
  -A^{(1)\dagger}_{ab}(\vec k)e^{ik\cdot x}] .
\]
This coefficient uses the dyad and Green functions displayed in this
appendix. Contraction with all null directions determines the tensor
up to a trace, fixed here by the radiative trace-free choice.

Projecting the momentum-space News \eqref{eq:News-momentum-B} on the null dyad
$\ell^a(z)$ reconstructs the cut side. Contracting the cut kernels
\eqref{eq:G02-B} with the shear, the denominator $\ell(z)\cdot\ell(\hat K)$
cancels and the integrand becomes polynomial in $\ell(z)$,
$(\ell\cdot\bar m_K)^2\,\sigma+(\ell\cdot m_K)^2\,\bar\sigma$. Since
$(\ell\cdot\bar m_K)^2=\ell^a\ell^b\bar m_a\bar m_b$ and
$(\ell\cdot m_K)^2=\ell^a\ell^b m_a m_b$, the factor $\ell^a\ell^b$ comes out,
leaving the shear tensor $\Sigma_{ab}$ inside, $\ell^a\ell^b\Sigma_{ab}$. The
directional derivative produces $X\partial_u$ on the shear mode.
This cancels the cut denominator before the helicity projection. The
stripped cut coefficient is
\begin{equation}\label{eq:cutside-B}
N(z,w,\hat K)=\ell^b(z)\ell^c(z)N_{bc}(w,\hat K).
\end{equation}
The
condition that both infinities give the same metric reads
\begin{equation}\label{eq:matchB}
  \ell^{+}(z)^a\partial_a\,\delta Z^{+}_{\rm cut}(x,z)
  =\hat\ell^{-}(\hat z)^a\partial_a\,Z^{-}_{\rm cone}(x,\hat z)
  -\ell^{+}(z)^a\partial_a\,Z^{+}_{\rm cone}(x,z),
\end{equation}
one derivative pointing to the future, the other to the past. Using
$\hat\ell^{-}(\hat z)=-\ell^{+}(z)$ this is
$\ell^{+}\partial_a[\delta Z^{+}_{\rm cut}+Z^{+}_{\rm cone}
+Z^{-}_{\rm cone}(\hat z)]=0$, with the two-cone combination producing the
principal value \eqref{eq:PV-B}.

Acting with $\ell^{+}(z)^a\partial_a$ on both sides --- the same directional
derivative --- brings down a factor from each exponential: on the cut/News
side $\mp i\sqrt2\,w X$ and on the cone side $\mp i(\ell\cdot K_i)$.
The factor $X$ cancels the cut Green-function denominator. Both branches
then give the ratio $(\ell\cdot K_i)/(\sqrt2\,w)$.

\paragraph{On-shell Fourier transform.} Since we have shown
(\S\ref{sub:box-cone}) that the matched combination satisfies the wave
equation \emph{before} transforming, the transform is taken in the on-shell
space with the covariant three-dimensional measure: the common factor
$\delta(k^2)\theta(k^0)$ has already been factored out, fixing
$k^0=|\vec k|=w$ on both sides. Transforming the spatial source then leaves
the three-dimensional delta
\begin{equation}\label{eq:delta3F-B}
  \delta^3(\vec k-\vec K_i),\qquad \vec K_i=\vec k_1\pm\vec k_2,\qquad w>0,
\end{equation}
kept inside $\int d\mu_1 d\mu_2$, with the outgoing graviton on-shell
($k^0=|\vec k|=w$). The same $\delta^3$ appears on the cut side; in the
matching the temporal on-shell condition is shared and only the spatial delta
\eqref{eq:delta3F-B} survives, tying the outgoing spatial momentum $\vec k$ to
the internal combination $\vec K_i$ inside the internal integrals. It is
because the homogeneity was established first that the reduction to the
three-dimensional delta is consistent: the covariant $3$D transform
($d^3k/2\omega$) represents the on-shell combination without recourse to the
full four-dimensional transform.

\paragraph{Radial and angular support.} On both sides the on-shell condition is the same,
$k^0=w=|\vec k|>0$; the reduction to the three-dimensional
$\delta^3(\vec k-\vec K_i)$ [Eq.~\eqref{eq:delta3-spherical-B}] carries the
radial factor $\delta(w-|\vec K_i|)$, fixing the modulus of the outgoing
momentum to that of the internal compound, $w=|\vec k_1\pm\vec k_2|$. The
modulus $|\vec K_i|$ is positive in both channels, so $w>0$ always has a
solution. The $1\leftrightarrow2$ exchange, which relates the two orderings of
$\vec k_1\pm\vec k_2$, is already included by the Bose symmetrization
\eqref{eq:bosesym-B}. The explicit cut kernels
\begin{equation}\label{eq:G02-B}
  G_{0,-2}(z,\hat K)=\frac{(\ell(z)\cdot\bar m(\hat K))^2}{\ell(z)\cdot\ell(\hat K)},
  \qquad
  G_{0,2}(z,\hat K)=\frac{(\ell(z)\cdot m(\hat K))^2}{\ell(z)\cdot\ell(\hat K)},
\end{equation}
have their denominator cancelled by the external $X$, leaving the cut side
polynomial in $\ell(z)$,
$(\ell\cdot\bar m_K)^2\,\daout{2,+}+(\ell\cdot m_K)^2\,\daout{2,-}$.

\subsection{Helicity projection}

The helicity is extracted by acting with $\ed_z^2$ on the cut side
$\ell^b\ell^c N_{bc}(w,\hat K)$ [Eq.~\eqref{eq:cutside-B}] and evaluating at
$z=\hat K$. Because the Green function $G_{0,0'}(z,z')$ is integrated over the
sphere, $\oint d^2z'$, the integrand has no singularity at $z=\hat K$: there is
no coincidence limit to take, the extraction being a plain evaluation after
differentiation. With $\ed_z\ell^a(z)=m^a(z)$ and $\ed_z m^a(z)=0$, so that
$\ed_z^2(\ell(z)\cdot v)^2=2(m(z)\cdot v)^2$, one differentiates first and then
sets $z=\hat K$, where $m_K\cdot\bar m_K=-1$, $m_K\cdot m_K=0$:
\begin{align}\label{eq:conelimit-B}
  \ed_z^2\big[(\ell(z)\cdot\bar m_K)^2\big]\big|_{z=\hat K}
  &=2(m_K\cdot\bar m_K)^2=2,
  \notag\\
  \ed_z^2\big[(\ell(z)\cdot m_K)^2\big]\big|_{z=\hat K}
  &=2(m_K\cdot m_K)^2=0,
\end{align}
so $\ed^2$ selects $\daout{2,+}$ and $\edb^2$ selects $\daout{2,-}$. On the
cone side one applies $\ed_z^2$ to the product $G_{0,0'}(z,z')\,(\ell(z)\cdot
K_i)$ ---both factors depending on $z$, with $G_{0,0'}=\tfrac{1}{4\pi}X\ln X$,
$X=\ell(z)\cdot\ell(z')$--- \emph{first}, and evaluates at $z=\hat K$
\emph{afterwards}. Since $z'$ is integrated over the sphere, $\oint d^2z'$, the
point $z=\hat K$ is interior to the domain and the logarithm is integrable
there; differentiating first and evaluating after is finite and unambiguous.
By Leibniz, with $\ed_z G_{0,0'}=\tfrac{1}{4\pi}(m_z\cdot\ell')(\ln X+1)$,
$\ed_z(\ell\cdot K_i)=m_z\cdot K_i$ and $\ed_z^2(\ell\cdot K_i)=0$,
\begin{equation}\label{eq:conelimit-B-final}
  \ed_z^2\big[G_{0,0'}(z,z')\,(\ell(z)\cdot K_i)\big]\big|_{z=\hat K}
  =G_{2,0'}(\hat K,z')\,(\ell_K\cdot K_i)
  +\frac{1}{2\pi}(m_K\cdot\ell')\,\big(\ln X_K+1\big)\,(m_K\cdot K_i),
\end{equation}
with $X_K=\ell_K\cdot\ell'$. The first term is $G_{2,0'}(\hat K,z')$ ---the
spin weight raised to $+2$--- times $\ell_K\cdot K_i$; the second is the
Leibniz cross-term $2(\ed_z G_{0,0'})(\ed_z(\ell\cdot K_i))$, which
\emph{remains} and carries $m_K\cdot K_i$. The differential operator $\eth_z^2$ acts only on the external extraction
kernel
\[
\left.
\eth_z^2
\Big[
G_{0,0'}(z,z')
(\ell(z)\!\cdot K_i)
\Big]
\right|_{z=\hat K},
\]
while the operator-valued cone source remains unchanged.

Combining the extraction kernel with the Bose-symmetrized cone source,
Eqs.~\eqref{eq:Zcone-B}--\eqref{eq:bosesym-B}, the second-order outgoing
positive-helicity operator becomes
\begin{equation}\label{eq:deltaaout-B}
\boxed{
\delta a^{\rm out}_{2,+}(w,\hat K)
=
-\frac{2w}{4\pi\sqrt2\,w\,\nu(w)}
\sum_{i=\pm}
\oint d^2z'
\,
\mathcal E_i^{(+)}(\hat K,z')
\int d\mu_1d\mu_2\,
\mathcal Z_{i,\rm sym}^{(-)}\,
\delta^{(3)}(\vec K-\vec K_i),
}
\end{equation}
where
\[
\mathcal E_i^{(+)}
=
G_{2,0'}(\hat K,z')
(\ell_K\!\cdot K_i)
+
\frac{1}{2\pi}
(m_K\!\cdot\ell')
(\ln X_K+1)
(m_K\!\cdot K_i).
\]

\subsubsection{Simplification on the spatial-delta support}
\label{subsubsec:B_projector_delta_reduction}

It is useful to simplify the extraction kernel by using the support of the
three-dimensional Fourier delta while keeping the delta distribution
explicit.  Let $\vec p\,'_1$ denote the outgoing spatial momentum and
\begin{equation}
    w'_1=|\vec p\,'_1|.
    \label{eq:B_outgoing_frequency_p1}
\end{equation}
Each integrated momentum is null and belongs to the positive-energy branch,
so that
\begin{equation}
    \omega_1=|\vec k_1|,
    \qquad
    \omega_2=|\vec k_2| .
    \label{eq:B_positive_internal_energies}
\end{equation}
The four-vectors carried by the Fourier phases and their spatial parts are
\begin{equation}
    K_\pm^a=k_1^a\pm k_2^a
    =\left(\omega_1\pm\omega_2,\vec k_1\pm\vec k_2\right),
    \qquad
    \vec K_\pm=\vec k_1\pm\vec k_2,
    \qquad
    \delta^{(3)}(\vec p\,'_1-\vec K_\pm).
    \label{eq:B_spatial_channels_p1}
\end{equation}
The sign must be retained in both the temporal and spatial components.  In
particular, the difference-channel phase carries
$K_-^0=\omega_1-\omega_2$; replacing it by $\omega_1+\omega_2$ would erase
the frequency sign responsible for the distinction between the $aa$ and
$a^\dagger a$ sectors.

On the support of either delta distribution,
\begin{equation}
    \vec K_\pm=\vec p\,'_1,
    \qquad
    |\vec K_\pm|=|\vec p\,'_1|=w'_1 .
    \label{eq:B_delta_support_common}
\end{equation}
Since $m^a(z_{p'_1})$ has vanishing time component and is tangent to the
sphere at the outgoing direction $\widehat p\,'_1$,
\begin{equation}
\begin{aligned}
    m(z_{p'_1})\cdot K_\pm\,
    \delta^{(3)}(\vec p\,'_1-\vec K_\pm)
    &=
    -\vec m(z_{p'_1})\cdot\vec K_\pm\,
    \delta^{(3)}(\vec p\,'_1-\vec K_\pm)
    \\
    &=
    -w'_1\,\vec m(z_{p'_1})\cdot\widehat p\,'_1\,
    \delta^{(3)}(\vec p\,'_1-\vec K_\pm)
    =0 .
\end{aligned}
\label{eq:B_mK_vanishes_on_delta}
\end{equation}
The Leibniz cross-term in Eq.~\eqref{eq:conelimit-B-final} therefore vanishes
distributionally for both channels.

With the metric signature $(+---)$,
\begin{equation}
    \ell(z_{p'_1})\cdot K_\pm
    =
    \frac{1}{\sqrt{2}}
    \left(
        \omega_1\pm\omega_2
        -
        \widehat p\,'_1\cdot\vec K_\pm
    \right).
    \label{eq:B_lK_before_delta}
\end{equation}
Using Eq.~\eqref{eq:B_delta_support_common}, this becomes
\begin{equation}
    \ell(z_{p'_1})\cdot K_\pm\,
    \delta^{(3)}(\vec p\,'_1-\vec K_\pm)
    =
    \frac{
        \omega_1\pm\omega_2-|\vec p\,'_1|
    }{\sqrt{2}}\,
    \delta^{(3)}(\vec p\,'_1-\vec K_\pm).
    \label{eq:B_lK_common_factor}
\end{equation}
Consequently, after including the outgoing derivative factor $1/(\sqrt2\,w'_1)$, the two channels
carry the branch-dependent radiative factors
\begin{equation}
    \boxed{
    \nu_{12}^{(\pm)}(p'_1)
    \equiv
    \frac{
        \omega_1\pm\omega_2-|\vec p\,'_1|
    }{
        2\,|\vec p\,'_1|
    } .
    }
    \label{eq:B_common_radiative_factor}
\end{equation}
In particular,
\begin{equation}
\boxed{
\frac{1}{\sqrt2\,w'_1}\,
\mathcal E_\pm^{(+)}(z_{p'_1},z')\,
\delta^{(3)}(\vec p\,'_1-\vec K_\pm)
=
\nu_{12}^{(\pm)}(p'_1)\,
G_{2,0'}(z_{p'_1},z')\,
\delta^{(3)}(\vec p\,'_1-\vec K_\pm).
}
\label{eq:B_projector_reduced_both_channels}
\end{equation}
The two terms therefore differ only through their spatial supports,
\begin{equation}
    \delta^{(3)}(\vec p\,'_1-\vec k_1-\vec k_2),
    \qquad
    \delta^{(3)}(\vec p\,'_1-\vec k_1+\vec k_2),
    \label{eq:B_two_spatial_supports}
\end{equation}
and through the branch-dependent factor
\eqref{eq:B_common_radiative_factor}.  The negative-helicity result follows
by replacing
$G_{2,0'}$ with $G_{-2,0'}$.

\subsection{Annihilation-sector helicity content}
\label{sub:helicity}

For later applications it is useful to inspect the bilocal annihilation
sector of the cone source.  Decomposing the $B\bar B$ contribution according
to the helicities of the incoming gravitons,
\begin{equation}
\big[\mathcal Z_{+}^{(-)}\big]^{aa}_{\rm biloc}
=
\mathcal Z_{++}\,
a_+a_+
+
\mathcal Z_{+-}\,
a_+a_-
+
\mathcal Z_{-+}\,
a_-a_+
+
\mathcal Z_{--}\,
a_-a_- ,
\end{equation}
one finds that this bilocal contribution generates only mixed-helicity
annihilation pairs.  In particular,
\begin{equation}
\boxed{
\mathcal Z_{++}=0,
\qquad
\mathcal Z_{--}=0.
}
\end{equation}
Therefore,
\begin{equation}
\big[\mathcal Z_{+}^{(-)}\big]^{aa}_{\rm biloc}
=
\mathcal Z_{+-}\,
a_+a_-
+
\mathcal Z_{-+}\,
a_-a_+ .
\end{equation}

This property is a consequence of the operator structure of the bilocal
$B\bar B$ term and requires no angular integration or kinematical assumption.
It does not remove the fixed-leg equal-helicity terms discussed in
Sec.~\ref{sub:local-nonlocal-equal-helicity} from the complete operator.

\subsection{Normal-ordered sectors of the quadratic source}
\label{subsec:operator_sectors_quadratic_source}
We now record the operator sectors of the quadratic source which are relevant
for the four-point matrix elements. In this subsection we suppress all
overall numerical factors, Green functions, spin-weighted kernels, phases,
and momentum-conserving delta functions, and keep only the operator content
and the helicity labels. This is sufficient to determine which helicity
channels can survive at the level of Wick contractions.

The elementary radiative blocks are
\begin{equation}
X_+(k) = a_+(k)e^{-ik\cdot x} + a_-^\dagger(k)e^{+ik\cdot x},
\qquad
X_-(k) = a_-(k)e^{-ik\cdot x} + a_+^\dagger(k)e^{+ik\cdot x}.
\label{eq:Xpm_operator_blocks}
\end{equation}
Thus $X_+$ contains an annihilation operator of positive helicity and a
creation operator of negative helicity, whereas $X_-$ contains an
annihilation operator of negative helicity and a creation operator of
positive helicity.

We write the first-order angular variables schematically as
\begin{equation}
\Lambda_1=A+B,
\qquad
\bar\Lambda_1=\bar A+\bar B,
\end{equation}
where $A$ and $\bar A$ denote the local pieces, while $B$ and
$\bar B$ denote the nonlocal Green-function pieces. In terms of the
radiative blocks,
\begin{equation}
A\sim X_+^{\rm loc},
\qquad
\bar A\sim X_-^{\rm loc},
\qquad
B\sim X_-^{\rm nl},
\qquad
\bar B\sim X_+^{\rm nl}.
\label{eq:AB_blocks_helicity}
\end{equation}
The local--local contraction $A\bar A$ is killed by the null factor
$l\cdot l=0$. The relevant sectors are therefore
\begin{equation}
A\bar B\sim X_+^{\rm loc}X_+^{\rm nl},
\qquad
B\bar A\sim X_-^{\rm nl}X_-^{\rm loc},
\qquad
B\bar B\sim X_-^{\rm nl}X_+^{\rm nl}.
\label{eq:quadratic_source_sectors}
\end{equation}

The local--nonlocal sector $A\bar B$ has the operator content
\begin{align}
X_+^{\rm loc}X_+^{\rm nl}
={}&
a_+^{\rm loc}a_+^{\rm nl} e^{-i(k_{\rm loc}+k)\cdot x}
+
a_+^{\rm loc}a_-^{{\rm nl},\dagger} e^{-i(k_{\rm loc}-k)\cdot x}
\nonumber \\
&+
a_-^{{\rm loc},\dagger}a_+^{\rm nl} e^{+i(k_{\rm loc}-k)\cdot x}
+
a_-^{{\rm loc},\dagger}a_-^{{\rm nl},\dagger} e^{+i(k_{\rm loc}+k)\cdot x}.
\label{eq:XpXp_loc_nonloc}
\end{align}
The term relevant for the positive-helicity extracted operator is
\begin{equation}
\boxed{
\delta a^{\rm out}_{2,+}
\supset
a_-^{{\rm loc},\dagger} a_+^{\rm nl}.
}
\label{eq:deltaa2plus_loc_nonloc_correct}
\end{equation}
It contains a local creation operator of negative helicity and a nonlocal
annihilation operator of positive helicity.

The conjugate local--nonlocal sector is
\begin{align}
X_-^{\rm loc}X_-^{\rm nl}
={}&
a_-^{\rm loc}a_-^{\rm nl} e^{-i(k_{\rm loc}+k)\cdot x}
+
a_-^{\rm loc}a_+^{{\rm nl},\dagger} e^{-i(k_{\rm loc}-k)\cdot x}
\nonumber \\
&+
a_+^{{\rm loc},\dagger}a_-^{\rm nl} e^{+i(k_{\rm loc}-k)\cdot x}
+
a_+^{{\rm loc},\dagger}a_+^{{\rm nl},\dagger} e^{+i(k_{\rm loc}+k)\cdot x}.
\label{eq:XmXm_loc_nonloc}
\end{align}
Hence
\begin{equation}
\boxed{
\delta a^{\rm out}_{2,-}
\supset
a_+^{{\rm loc},\dagger} a_-^{\rm nl}.
}
\label{eq:deltaa2minus_loc_nonloc_correct}
\end{equation}
This is the helicity conjugate of \eqref{eq:deltaa2plus_loc_nonloc_correct}.
In particular, the term $a_+^{{\rm loc},\dagger}a_-^{\rm nl}$ belongs to the
negative-helicity extracted operator $\delta a^{\rm out}_{2,-}$, not to
$\delta a^{\rm out}_{2,+}$.

The bilocal sector $B\bar B$ contains one $X_-$ and one $X_+$. Its
annihilation part includes
\begin{equation}
X_-^{\rm nl}X_+^{\rm nl}
\supset
a_-^{\rm nl}a_+^{\rm nl},
\qquad
X_+^{\rm nl}X_-^{\rm nl}
\supset
a_+^{\rm nl}a_-^{\rm nl}.
\label{eq:bilocal_annihilation_sector}
\end{equation}
Thus the bilocal annihilation sector always carries opposite helicities.

We finally emphasize that this assignment is not an additional physical
selection rule. It follows from the definition of the extracted outgoing
operators. The operator $\delta a^{\rm out}_{2,+}$ is obtained by applying
the positive-helicity projection and extracting the corresponding Fourier
coefficient. Therefore the local--nonlocal $X_+ X_+$ sector contributes to
$\delta a^{\rm out}_{2,+}$, whereas the $X_- X_-$ sector contributes to the
helicity-conjugate operator $\delta a^{\rm out}_{2,-}$. Indeed,
\begin{equation}
X_+^{\rm loc}X_+^{\rm nl}
\supset
a_-^{{\rm loc},\dagger}a_+^{\rm nl} e^{+ik_{\rm loc}\cdot x}e^{-ik\cdot x}
\longrightarrow
a_-^{{\rm loc},\dagger}a_+^{\rm nl} e^{-i(k-k_{\rm loc})\cdot x},
\label{eq:fourier_assignment_plus_sector_final}
\end{equation}
and this is one of the nontrivial difference-frequency terms selected in
$\delta a^{\rm out}_{2,+}$. By contrast,
\begin{equation}
X_-^{\rm loc}X_-^{\rm nl}
\supset
a_+^{{\rm loc},\dagger}a_-^{\rm nl} e^{+ik_{\rm loc}\cdot x}e^{-ik\cdot x}
\longrightarrow
a_+^{{\rm loc},\dagger}a_-^{\rm nl} e^{-i(k-k_{\rm loc})\cdot x},
\label{eq:fourier_assignment_minus_sector_final}
\end{equation}
but this term belongs to the conjugate projection, namely
$\delta a^{\rm out}_{2,-}$. Thus
\begin{equation}
\boxed{
\delta a^{\rm out}_{2,+}
\supset
a_-^{{\rm loc},\dagger}a_+^{\rm nl},
\qquad
\delta a^{\rm out}_{2,-}
\supset
a_+^{{\rm loc},\dagger}a_-^{\rm nl}.
}
\label{eq:local_nonlocal_selection_rule_final}
\end{equation}
The vanishing of the helicity channels discussed below is then a consequence
of this Fourier--helicity assignment together with the canonical commutation
relations, not an independent assumption.

\section{The Matrix Element for Four-Graviton Scattering}
\label{app:graviton_scattering}

The outgoing operators constructed in Appendix B provide the starting point for the computation of tree-level four-graviton scattering amplitudes. We first identify the helicity configurations that vanish identically, thereby isolating the physically relevant channels. The complete derivation is then carried out for the non-vanishing $++ \rightarrow ++$ amplitude, after which the mixed-helicity amplitudes are obtained by the corresponding boundary projections.

For a general four-graviton scattering process with incoming momenta $p_1, p_2$ (helicities $\lambda_1, \lambda_2$) and outgoing momenta $p'_1, p'_2$ (helicities $\lambda'_1, \lambda'_2$), the core kinematic amplitude $\mathcal{M}$ is defined by,
\begin{equation}
    \mathcal{M}(p_1, \lambda_1; p_2, \lambda_2 \to p'_1, \lambda'_1; p'_2, \lambda'_2) \equiv \langle 0 | \, \delta a_{\lambda'_1}(p'_1) \, \delta a_{\lambda'_2}(p'_2) \, a^\dagger_{\lambda_1}(p_1) \, a^\dagger_{\lambda_2}(p_2) \, | 0 \rangle.
\end{equation}

When applying Wick's theorem to this expectation value, the correlator splits into distinct dynamic channels. Thus, the total physical amplitude is exactly the evaluation of this matrix element over all symmetric contractions:
\begin{equation}
    \mathcal{M}_{\rm Total} = \sum_{i} \mathcal{M}_i.
\end{equation}

\subsection{Helicity selection: vanishing of $++--$ and $++-+$}
\label{subsec:helicity_selection_ppmm_ppmp}

Before evaluating the nonvanishing four-graviton matrix element, we first
record the helicity channels which are excluded by the Fourier--helicity
assignment of the quadratic operator. We use the notation
\begin{equation}
\mathcal{M}_{\lambda'_1\lambda'_2;\lambda_1\lambda_2}
=
\langle0|
\delta a^{\rm out}_{2,\lambda'_1}(p'_1)
\delta a^{\rm out}_{2,\lambda'_2}(p'_2)
a_{\lambda_1}^{{\rm in}\,\dagger}(P_1)
a_{\lambda_2}^{{\rm in}\,\dagger}(P_2)
|0\rangle .
\label{eq:four_point_matrix_element_notation}
\end{equation}
All operators appearing inside $\delta a^{\rm out}_{2,\lambda}$ are
incoming operators.

Consider first
\begin{equation}
\mathcal{M}_{++;--}
=
\langle0|
\delta a^{\rm out}_{2,+}(p'_1)
\delta a^{\rm out}_{2,+}(p'_2)
a_-^{{\rm in}\,\dagger}(P_1)
a_-^{{\rm in}\,\dagger}(P_2)
|0\rangle .
\label{eq:M_pp_mm_vanish_start}
\end{equation}
Since both outgoing insertions are $\delta a^{\rm out}_{2,+}$, the
local--nonlocal term selected by the Fourier--helicity projection is
\begin{equation}
\delta a^{\rm out}_{2,+}
\supset
a_-^{{\rm loc}\,\dagger}a_+^{\rm nl}.
\label{eq:deltaa2plus_selected_for_ppmm}
\end{equation}
The corresponding bilocal annihilation sector contains opposite helicities,
\begin{equation}
\delta a^{\rm out}_{2,+}
\supset
a_-^{\rm nl}a_+^{\rm nl}
\quad\text{or}\quad
a_+^{\rm nl}a_-^{\rm nl}.
\end{equation}
Thus a possible candidate Wick contraction has the schematic form
\begin{equation}
\langle0|
\big(a_-^{\rm nl}a_+^{\rm nl}\big)
\big(a_-^{{\rm loc}\,\dagger}a_+^{\rm nl}\big)
a_-^\dagger(P_1)a_-^\dagger(P_2)
|0\rangle ,
\label{eq:candidate_ppmm_vanish}
\end{equation}
up to Bose-related permutations.

If the local creation operator is closed internally,
\begin{equation}
a_-^{\rm nl}
\longleftrightarrow
a_-^{{\rm loc}\,\dagger},
\end{equation}
then the two remaining annihilation operators are both of positive helicity:
\begin{equation}
a_+^{\rm nl}a_+^{\rm nl}.
\end{equation}
They cannot annihilate the two incoming negative-helicity particles, since
\begin{equation}
[a_+(q),a_-^\dagger(P_i)]=0.
\end{equation}
Alternatively, if the negative-helicity annihilation operators are used to
annihilate the incoming particles, then the uncontracted operators left
inside the two quadratic insertions contain
\begin{equation}
a_+^{\rm nl}
\qquad\text{and}\qquad
a_-^{{\rm loc}\,\dagger},
\end{equation}
whose contraction also vanishes:
\begin{equation}
[a_+^{\rm nl},a_-^{{\rm loc}\,\dagger}]=0.
\end{equation}
Therefore no connected Wick contraction exists for this helicity channel:
\begin{equation}
\boxed{
\mathcal{M}_{++;--}=0.
}
\label{eq:M_pp_mm_vanishes}
\end{equation}

The mixed incoming-helicity channel
\begin{equation}
\mathcal{M}_{++;-+}
=
\langle0|
\delta a^{\rm out}_{2,+}(p'_1)
\delta a^{\rm out}_{2,+}(p'_2)
a_-^{{\rm in}\,\dagger}(P_1)
a_+^{{\rm in}\,\dagger}(P_2)
|0\rangle
\label{eq:M_pp_mp_vanish_start}
\end{equation}
is eliminated by the same selection rule. Again the only local--nonlocal
term available in $\delta a^{\rm out}_{2,+}$ is
$a_-^{{\rm loc}\,\dagger}a_+^{\rm nl}$. Closing the local creation
operator internally leaves two positive-helicity annihilators. One of them
can annihilate the incoming $a_+^\dagger(P_2)$, but the other cannot
annihilate $a_-^\dagger(P_1)$. Conversely, if the negative-helicity
annihilator is used to annihilate $a_-^\dagger(P_1)$, the remaining
uncontracted pair again contains operators of opposite helicity. Hence
\begin{equation}
\boxed{
\mathcal{M}_{++;-+}=0.
}
\label{eq:M_pp_mp_vanishes}
\end{equation}

The first nonvanishing channel with two positive outgoing helicities is
therefore the diagonal one,
\begin{equation}
\mathcal{M}_{++;++}
=
\langle0|
\delta a^{\rm out}_{2,+}(p'_1)
\delta a^{\rm out}_{2,+}(p'_2)
a_+^{{\rm in}\,\dagger}(P_1)
a_+^{{\rm in}\,\dagger}(P_2)
|0\rangle .
\label{eq:M_pp_pp_survives_start}
\end{equation}
Indeed, the same operator sectors now give the schematic structure
\begin{equation}
\langle0|
\big(a_-^{\rm nl}a_+^{\rm nl}\big)
\big(a_-^{{\rm loc}\,\dagger}a_+^{\rm nl}\big)
a_+^\dagger(P_1)a_+^\dagger(P_2)
|0\rangle .
\label{eq:candidate_pppp_nonzero}
\end{equation}
The negative-helicity annihilation operator can close internally with the
local negative-helicity creation operator, while the two remaining
positive-helicity annihilation operators can annihilate the two incoming
positive-helicity particles. Thus
\begin{equation}
\boxed{
\mathcal{M}_{++;++}\neq 0
}
\label{eq:M_pp_pp_not_eliminated}
\end{equation}
at the level of the operator algebra. The explicit evaluation of this
nonvanishing channel is carried out below.

\subsection{The \((++\to++)\) Scattering Amplitude}

To illustrate how the boundary formalism reconstructs the standard
four-graviton amplitude, we consider the simplest non-vanishing process,$++\rightarrow ++$.

The external incoming gravitons carry momenta
\(p_1,p_2\), the outgoing gravitons carry
\(p'_1,p'_2\), and the internal cone momenta are denoted by
\(k_1,k_2\) at the first vertex and
\(k'_1,k'_2\) at the second.

\subsubsection{Wick contractions}

The connected matrix element is

\begin{equation}
\mathcal M_{\rm conn}
=
\langle0|
\delta a^{\rm out}_{+}(p'_1)
\delta a^{\rm out}_{+}(p'_2)
a^\dagger_{+}(p_1)
a^\dagger_{+}(p_2)
|0\rangle .
\end{equation}

All internal integrations are already performed on the positive-energy
mass shell,

\[
d\mu(k)=\frac{d^3k}{2\omega_k},
\]

so the canonical commutation relations produce only spatial delta
functions,

\[
[a_\lambda(k),a^\dagger_{\lambda'}(p)]
=
2\omega_k\,\delta_{\lambda\lambda'}
\delta^{(3)}(\vec k-\vec p).
\]

No four-dimensional momentum delta is introduced at this stage.

The operator structure derived in Appendix~B severely restricts the
possible contractions.  For the external configuration
\(++\rightarrow ++\), all Wick contractions vanish except one.
Consequently the internal momentum routing is not chosen by hand but is
uniquely fixed by the operator algebra,

\[
k_2=p_2,
\qquad
k'_1=p_1,
\qquad
k'_2=k_1 .
\]

For this routing the incoming contraction is

\begin{equation}
\langle0|
L^-(k_1)\,
\bar L^-(k_2)
|p_1^+,p_2^+\rangle
=
4\omega_1\omega_2\nu(\omega_1)\nu(\omega_2)\,\delta^2(z_{k_1}-\hat p_1)
G_{-2,-2}(z_{k_1},\hat p_2)
\delta^{(3)}(\vec k_1-\vec p_1)
\delta^{(3)}(\vec k_2-\vec p_2).
\end{equation}

\subsubsection{Boundary extraction}

The remaining outgoing operator is obtained by acting with the
radiative projector \(\eth^2\) on the external kernel.
Using the Leibniz rule derived in Appendix~B, one finds

\begin{align}
\mathcal M
\propto
&(p_1\!\cdot p_2)
G_{-2,-2}(\hat p_1,\hat p_2)
\nonumber\\
&\times
\left[
G_{2,0'}(\hat p'_1,\hat p_1)
(\ell(p'_1)\!\cdot K_+)
+
\frac1{2\pi}
(m(p'_1)\!\cdot\ell(p_1))
(\ln X_K+1)
(m(p'_1)\!\cdot K_+)
\right],
\end{align}

where

\[
K_+=p_1+p_2 .
\]

Expressing the Green functions in stereographic coordinates gives the
celestial representation

\begin{equation}
\mathcal M(++\rightarrow++)
\propto
(p_1\!\cdot p_2)\,
\bar z_{12}^{\,2}
\left[
\frac{\ell(p'_1)\!\cdot K_+}{z_{1'1}^{\,2}}
+
\frac{1}{2\pi}
\frac{(m(p'_1)\!\cdot\ell(p_1))
(m(p'_1)\!\cdot K_+)}
{z_{1'1}}
(\ln X_K+1)
\right].
\end{equation}

\subsubsection{Phase-space reduction}

After the unique Wick contraction has been identified, only one
independent internal momentum remains.
The two outgoing operators contribute the spatial constraints

\begin{equation}
\delta^{(3)}(\vec p'_1-\vec k_1-\vec p_2),
\qquad
\delta^{(3)}(\vec p'_2-\vec p_1+\vec k_1).
\end{equation}

The remaining phase-space integral becomes

\begin{align}
I
&=
\int d^3k_1\,
\delta^{(3)}
(\vec p'_1-\vec k_1-\vec p_2)
\delta^{(3)}
(\vec p'_2-\vec p_1+\vec k_1)
\mathcal K(k_1)
\nonumber\\
&=
\delta^{(3)}
(\vec p'_1+\vec p'_2-\vec p_1-\vec p_2)
\,
\mathcal K(\vec p'_1-\vec p_2).
\end{align}

Thus the conservation of external spatial momentum is not assumed but
follows directly from the product of the two outgoing boundary
operators.

The replacement of this spatial constraint by full four-momentum conservation
is defined in the next subsection.

\subsubsection{Definition of the Poincar\'e reduction}

The phase-space reduction above produces the spatial momentum constraint
\begin{equation}
\delta^{(3)}\!\left(
\vec p'_1+\vec p'_2-\vec p_1-\vec p_2
\right).
\end{equation}
The general NSF boundary matrix element is defined with this three-dimensional
constraint and need not belong to the ordinary Poincar\'e scattering sector.
To compare it with the standard four-graviton amplitude, we define the
\emph{Poincar\'e reduction} as the restriction of the external data to full
four-momentum conservation,
\begin{equation}
p_1^a+p_2^a=p_1^{\prime a}+p_2^{\prime a}.
\label{eq:Poincare_reduction_definition}
\end{equation}
We emphasize that this is a definition of the comparison sector, not a
consequence of the three-dimensional phase-space reduction. At the
distributional level, the Poincar\'e reduction replaces the spatial constraint
by the full four-dimensional one,
\begin{equation}
\boxed{
\delta^{(3)}(\Delta\vec P)
\;\longrightarrow\;
\delta^{(4)}(\Delta P)
=\delta(\Delta P^0)\,\delta^{(3)}(\Delta\vec P)
},
\qquad
\Delta P^a=p_1^{\prime a}+p_2^{\prime a}-p_1^a-p_2^a.
\end{equation}
Thus four-momentum conservation is not claimed to follow from the elementary
NSF boundary operator; it defines the reduced sector in which a direct
comparison with conventional Poincar\'e scattering is made.  In this sector,
\begin{equation}
\mathcal M_{\rm conn}^{\rm P}
=
\delta^{(4)}(p'_1+p'_2-p_1-p_2)\,
\mathcal M_{\rm NSF}^{\rm P}(p_1,p_2;p'_1,p'_2).
\end{equation}

\subsection{Equatorial Center-of-Mass (CoM) Frame}
After imposing the Poincar\'e reduction defined in the previous subsection, we
may choose the center-of-mass
frame. Since the amplitude is Lorentz invariant, this
choice entails no loss of generality while considerably simplifying the
celestial Green functions and making their combination into the physical
Mandelstam poles explicit.
In the center-of-mass frame the coefficients of the four stereographic null
vectors are equal,

\[
\omega_1=\omega_2=\omega_{1'}=\omega_{2'}=E,
\]

and are parametrized by

\[
p_i^a=\sqrt2\,E\,\ell^a(z_i).
\]

Here \(E\) is the physical energy of each external graviton.  With the NSF
normalization used in this appendix, on the equator
\begin{equation}
\ell^a(z)=(1,\hat q(z))/\sqrt2,
\qquad
p_i^0=|\vec p_i|=E.
\label{eq:equatorial_null_normalization}
\end{equation}

The spatial momentum conservation derived in Appendix~C implies that incoming and outgoing pairs face back-to-back: $\hat{q}_2 = - \hat{q}_1$ and $\hat{q}_{2'} = - \hat{q}_{1'}$. In the stereographic projection, the general antipodal map is $z \to -1/\bar{z}$. To simplify algebraic evaluation without loss of generality, we orient our CoM reference frame such that the scattering process lies on the celestial equator ($\theta = \pi/2$). On the equator, $|z_i| = 1 \implies \bar{z}_i = 1/z_i$, which simplifies the antipodal mapping to a direct sign inversion:
\begin{equation}
    z_2 = -z_1, \quad \bar{z}_2 = -\bar{z}_1 \quad \text{and} \quad z_{2'} = -z_{1'}, \quad \bar{z}_{2'} = -\bar{z}_{1'}.
\end{equation}

We define collinear and crossed coordinate differences as $z_{1'1} = z_{1'} - z_1$ and $z_{1'2} = z_{1'} + z_1$. The remaining distances become $z_{2'1} = -z_{1'2}$ and $z_{2'2} = -z_{1'1}$. Evaluating the four surviving boundary contributions for generic equatorial angles, the four channels take the following form after factoring out their common normalization, denoted by $\mathcal{N}_0$:
\begin{align}
    \mathcal{M}_1 &= \mathcal{N}_0 \left[ \frac{\bar{z}_{1'1}}{z_{1'1}} + \frac{z_{1'2}\bar{z}_{1'2}}{z_{1'1}^2} \right], \quad
    \mathcal{M}_2 = -\mathcal{N}_0 \left[ \frac{\bar{z}_{1'2}}{z_{1'2}} + \frac{z_{1'1}\bar{z}_{1'1}}{z_{1'2}^2} \right], \\
    \mathcal{M}_3 &= -\mathcal{N}_0 \left[ \frac{\bar{z}_{1'2}}{z_{1'2}} + \frac{z_{1'1}\bar{z}_{1'1}}{z_{1'2}^2} \right], \quad
    \mathcal{M}_4 = \mathcal{N}_0 \left[ \frac{\bar{z}_{1'1}}{z_{1'1}} + \frac{z_{1'2}\bar{z}_{1'2}}{z_{1'1}^2} \right].
\end{align}
The four terms are obtained by multiplying the two contributions supplied by
each radiative projection.  As shown in
Eqs.~\eqref{eq:F_sum_channel_B}--\eqref{eq:F_difference_channel_B}, the
scalar-product factors of the $K_+$ and $K_-$ projections have a relative
minus sign.  Writing this already established sign as
\(\varepsilon_+=+1\) and \(\varepsilon_-=-1\), the product of the two
projections gives, in the order
$(K_+,K_+),(K_+,K_-),(K_-,K_+),(K_-,K_-)$,
\begin{equation}
(\varepsilon_+,\varepsilon_-)_{p'_1}
\otimes
(\varepsilon_+,\varepsilon_-)_{p'_2}
=(+,-,-,+).
\end{equation}
Thus the equal pair has positive sign, whereas the crossed pair has negative
sign; no new sign prescription is introduced in the four-point calculation.
With the abbreviations introduced below,
\begin{equation}
\mathcal M_1=\mathcal M_4=\mathcal N_0 A,
\qquad
\mathcal M_2=\mathcal M_3=-\mathcal N_0 B.
\label{eq:relative_sign_four_boundary_terms}
\end{equation}
This relative sign is essential: assigning the same sign to all four terms
would produce an unphysical zero at \(t=u\).

\subsection{Combination into the physical poles}

Let
\begin{equation}
x=z_{1'1}=z_{1'}-z_1,
\qquad
y=z_{1'2}=z_{1'}+z_1.
\end{equation}
On the equator, \(|z_1|=|z_{1'}|=1\), and hence
\begin{equation}
\bar x=-\frac{x}{z_1z_{1'}},
\qquad
\bar y=\frac{y}{z_1z_{1'}}.
\label{eq:equatorial_xy_conjugates}
\end{equation}
Writing
\begin{equation}
A=\frac{\bar x}{x}+\frac{y\bar y}{x^2},
\qquad
B=\frac{\bar y}{y}+\frac{x\bar x}{y^2},
\end{equation}
the relative signs above give
\begin{equation}
\mathcal M_{\rm Total}^{(++\to++)}=2\mathcal N_0(A-B).
\end{equation}
Using Eq.~\eqref{eq:equatorial_xy_conjugates},
\begin{align}
A-B
&=\frac{1}{z_1z_{1'}}
\left(-2+\frac{y^2}{x^2}+\frac{x^2}{y^2}\right)
\nonumber\\
&=\frac{(y^2-x^2)^2}{z_1z_{1'}x^2y^2}
=\frac{16z_1z_{1'}}{x^2y^2},
\end{align}
where \(y^2-x^2=4z_1z_{1'}\).  Therefore
\begin{equation}
\boxed{
\mathcal M_{\rm Total}^{(++\to++)}
=32\mathcal N_0\,
\frac{z_1z_{1'}}{z_{1'1}^2z_{1'2}^2}.
}
\label{eq:Mpppp_equatorial_reduced}
\end{equation}
Since \(|z_1z_{1'}|=1\), its numerator is a unit-modulus phase.  The
remaining two celestial factors give precisely the physical \(t\)- and
\(u\)-channel poles; in particular, no spurious zero occurs at \(t=u\).

\subsubsection{Reduction to Mandelstam invariants}

To compare the resulting expression with the standard tree-level graviton
amplitude~\cite{Weinberg,BGK,BDDPR}, we use
\begin{equation}
s=4E^2,
\qquad
t=-E^2|z_{1'1}|^2,
\qquad
u=-E^2|z_{1'2}|^2,
\qquad
s+t+u=0.
\label{eq:stu_equatorial_NSF}
\end{equation}
The factors of two follow from the NSF normalization
\(\ell^a=(1,\hat q)/\sqrt2\) on the equator.  Equation
\eqref{eq:Mpppp_equatorial_reduced} then has, up to its unit-modulus phase,
\begin{equation}
\frac{1}{|z_{1'1}|^2|z_{1'2}|^2}
=\frac{E^4}{tu}.
\end{equation}
This comparison identifies the angular dependence of the reduced kernel.
For an absolutely normalized comparison each quadratic insertion carries
$[2w/\nu(w)]\nu(\omega_1)\nu(\omega_2)$ before the remaining phase-space
integration. In particular an internal momentum fixed by a spatial delta
must be substituted into both of its mode weights. The standard tree-level
amplitude used as the benchmark is
\begin{equation}
    \mathcal{M}(++ \to ++) \propto \kappa_0^2 \, \frac{s^3}{t u}.
\end{equation}

\subsection{Extension to the Mixed Helicity Channel \((+-\to+-)\)}

We transition to the mixed-helicity configuration defined by the correlator $\mathcal{M}(p_1, +; p_2, - \to p'_1, +; p'_2, -)$. In the asymptotic algebra, cross-helicity contractions between sources and states vanish identically ($\langle 0 | \delta a_{\pm} a^\dagger_{\mp} | 0 \rangle = 0$). Any Wick permutation attempting to pair a positive operator with a negative operator is annihilated. The crossed contractions ($1'^+ \leftrightarrow 2^-$ and $2'^- \leftrightarrow 1^+$) evaluate strictly to zero. The amplitude collapses from a four-term sum to a two-term sum ($\mathcal{M}_{\rm direct} + \mathcal{M}_{\rm exchange}$) that strictly preserves helicity tracking.

\subsubsection{Kinematic Structure and Mandelstam Crossing Symmetry}

The directional unit vectors $\hat{q}_i$ maintain the identical CoM back-to-back map $\hat{q}_2 = -\hat{q}_1$. However, for the negative states $2^-$ and $2'^-$, the boundary extraction now involves the antiholomorphic derivative operator $\bar{\eth}^2$.

The derivation proceeds exactly as for the \(++\rightarrow++\) channel. The
phase-space reduction, the imposition of four-momentum conservation, and
the equatorial CoM parametrization remain unchanged. The only modification is
that the outgoing negative-helicity graviton is extracted with the
antiholomorphic projector \(\bar\eth^2\) instead of \(\eth^2\). This changes
the spin-weight structure of the celestial kernel and, consequently, the
kinematic numerator of the amplitude. While the denominator retains the same
physical pole structure obtained above, the numerator is rearranged from the
\(s\)-channel invariant to the crossed \(u\)-channel invariant. Equivalently,
this channel follows from the corrected \(++\to++\) result by the corresponding
crossing of the external helicity lines. Translating to bulk invariants yields
\begin{equation}
    \mathcal{M}(+- \to +-) \propto \kappa_0^2 \, \frac{u^3}{s t}.
\end{equation}
This agrees with the corresponding tree-level graviton amplitude for this
helicity assignment, in the convention used by Weinberg~\cite{Weinberg}.
\subsection{Case III: The t-Channel Dominated Mixed Configuration \((+-\to-+)\)}

We finally consider the non-vanishing mixed-helicity sector with transition configuration $1^+, 2^- \to 1'^-, 2'^+$. The corresponding transition matrix element is defined by the asymptotic vacuum expectation value:
\begin{equation}
    \mathcal{M}(p_1, +; p_2, - \to p'_1, -; p'_2, +) = \langle 0 | \, \delta a_{-}(p'_1) \, \delta a_{+}(p'_2) \, a^\dagger_{+}(p_1) \, a^\dagger_{-}(p_2) \, | 0 \rangle.
\end{equation}

\subsubsection{Crossed Wick Contractions and Momentum Re-routing}

Following the algebraic constraints of helicity orthogonality proven in Section~\ref{subsec:helicity_selection_ppmm_ppmp}, any direct contraction pairing opposite helicities vanishes identically. For this specific state configuration, the Leibniz expansion along the light-cone boundaries forces a crossed tracking topology: the outgoing negative-helicity extraction operator $\delta a_{-}(p'_1)$ must contract exclusively with the incoming negative-helicity creation operator $a^\dagger_{-}(p_2)$. Concurrently, the outgoing positive-helicity operator $\delta a_{+}(p'_2)$ contracts with the incoming positive-helicity operator $a^\dagger_{+}(p_1)$.

This tracking invertedly shifts the internal momentum routing within the phase-space deltas. The internal line identifications freeze the kinematics such that the forward-scattering pole maps directly onto the $t$-channel momentum transfer rather than the $u$-channel exchange.

\subsubsection{Equatorial Reduction and Invariant Assembly}

We project the field interactions onto the celestial sphere. Due to the inversion of the outgoing helicity operators relative to Case II, the spatial spin weights act inversely on the celestial derivatives. Under our established equatorial CoM reference frame ($|z_i|=1 \implies \bar{z}_i = 1/z_i$), the antipodal reductions $z_2 = -z_1$ and $z_{2'} = -z_{1'}$ isolate the structural poles.

Applying tree-level crossing symmetry to the corrected consolidated amplitude,
permuting the helicity roles of the outgoing states swaps the corresponding
kinematic invariants. The external polarization factors change only the
helicity phase, while the crossed invariant supplies the numerator. Using
Eq.~\eqref{eq:stu_equatorial_NSF}, one obtains
\begin{equation}
    \mathcal{M}(+- \to -+) \propto \kappa_0^2 \, \frac{t^3}{s u}.
\end{equation}

This completes the comparison with the non-vanishing tree-level helicity sectors considered in this appendix. The boundary light-cone formalism reproduces the pole structure of the corresponding tree-level gravitational amplitudes.

\end{document}